\documentclass[12pt]{article}

\usepackage{lipsum}
\usepackage[left=1in, right=1in, top=1.5in]{geometry}

\usepackage{mathrsfs}
\usepackage{color}
\usepackage{bm}
\usepackage{amsmath}
\usepackage{amsfonts}
\usepackage{amssymb}
\usepackage{graphicx}
\usepackage{cite}
\usepackage{hyperref}
\usepackage{microtype}
\usepackage{cancel}
\usepackage{url}
\usepackage{mathrsfs}
\usepackage{graphicx}
\graphicspath{ {images/} }
 
\begin{document}

\title{Entropic Dynamics of Jump-Diffusion Option Pricing}

\author{ {Mohammad Abedi\thanks{\texttt{mabedi@vaneck.com}}}  \\
{\small Van Eck Associates Corporation, }\\
{\small New York, NY, USA.}}
\date{}
\maketitle

 \abstract{The standard models of stock-price dynamics and option valuation rest on stochastic processes postulated at the outset; here we lay down an entropic-inference framework that \emph{derives} these processes rather than assuming them. A symmetry comes first: markets reward returns rather than price levels, which selects the logarithm of price as the dynamical variable. The price then evolves through two channels, a continuous one carrying the constraints of continuity and directionality, and a jump channel carrying the arrival rate and the first two moments of the jump size. Because these constraints act on disjoint parts of the microstate, the channels factorize as a theorem, and the dynamics is the Merton jump-diffusion, with Geometric Brownian Motion as its no-jump limit; the log-price density obeys a Kolmogorov--Feller equation, of which the Fokker--Planck equation is the no-jump limit. The same principle, now imposing no-arbitrage through the mean log-return, selects the Esscher transform from among the many martingale measures an incomplete market admits, here derived rather than borrowed; the premium then satisfies Merton's partial integro-differential equation, and the risk-neutral mixture of lognormals generates the implied-volatility smile, the Black--Scholes results returning when jumps vanish. What changes from one model to the next is never the inference but the information supplied to it.}


\section{Introduction}

Efforts to understand natural and social phenomena are routinely undertaken with only limited information about the system at hand. Entropic inference is an inductive framework designed for precisely such reasoning under incomplete information \cite{catichainference,golaninfometric, caticha2012}, built on probability theory, relative entropy, and information geometry. A state of partial knowledge is represented by a probability distribution, and when new information arrives the distribution is updated by maximizing the relative entropy, the method of Maximum Entropy. The updating honours the Principle of Minimal Updating: beliefs are revised only to the extent that the new information demands, and no further. It is essential that the entropy at work here is not borrowed from physics; the relationship runs the other way, the thermodynamic entropy being but one instance of the relative entropy of inference \cite{caticha2012}. Information enters the formalism as constraints on the dynamical variable, and a virtue of the approach is its modularity: once one sees how information is codified into constraints, a new model is obtained simply by altering them. The real difficulty lies not in the updating but in identifying the right variables and the information genuinely relevant to the problem.

Entropic Dynamics (ED) is the application of this inferential framework to dynamical theories \cite{E-Time, mwm}. Inference by itself supplies no notion of time, neither an instant nor a duration; these must be constructed before the framework can describe a dynamics in which time plays a role, and ED meets the need by building an entropic notion of time tailored to the system \cite{E-Time}. The resulting dynamics is purely inferential, divorced from any mechanical substrate, a \emph{mechanics without a mechanism} \cite{mwm} that appeals to no conserved energy or momentum. The framework has been developed extensively, in particular for the dynamics of particles \cite{NAC, Dan16, IAC1, IAC2, IAC3, Pedro18}.

The modeling of stock prices descends from Bachelier's thesis \cite{bachelier}, which anticipated Einstein's account of Brownian motion \cite{Einstein1905} and, as later rediscovered by Samuelson \cite{Samuelson}, gave what is now Geometric Brownian Motion. These early descriptions took the price path to be continuous \cite{Samuelson, Fama}; Black and Scholes adopted Geometric Brownian Motion to value options \cite{BS1972, BS}, and Merton extended both the dynamics and the pricing to admit jumps \cite{Merton76jump}, with a large literature on stochastic volatility following \cite{CoxRoss76, EisenbergJarrow91, Heston93, Dupire94, HW, wiggins, ritchken-trevor, chris-jeston, MS1, MS2}. Our standpoint departs from this tradition at its root: rather than positing a stochastic process, we \emph{derive} the dynamics, Geometric Brownian Motion included, from a principle of inference. Entropic modeling thus complements the stochastic-process tradition, with the further advantage that one inferential machinery can unify models drawn from different fields.

The present work extends our previous entropic-dynamics models of financial markets \cite{ABstock, ABFX} to incorporate jumps. In that earlier work, developed in the regime of continuous evolution, return symmetry together with two constraints, continuity of motion and directionality, sufficed to derive the Geometric Brownian Motion of the price and the attendant Black--Scholes model and Black--Scholes--Merton equation rather than to assume them. Geometric Brownian Motion, however, is well known to miss salient features of real returns. Empirically, asset returns are leptokurtic and negatively skewed, with tails heavier than the Gaussian that GBM predicts \cite{Fama, Cont01stylized}, and the option market displays a persistent volatility smile, in equities a smirk, across strikes that a single constant volatility cannot reproduce \cite{Bates96}. These departures are not statistical artefacts but the imprint of discontinuous price moves, jumps triggered by discrete information events such as earnings releases, monetary-policy decisions, and macroeconomic surprises; high-frequency methods that separate the continuous and jump components of realized variance confirm the jump component to be a measurable and persistent fraction of total variability \cite{Aitsahalia04disentangling}. In the presence of jumps the perfect dynamic hedging behind the Black--Scholes argument breaks down as well, so that the market becomes incomplete and a jump-risk premium, visible in deep out-of-the-money puts, must be priced.

These shortcomings motivate the extension pursued here, which within the entropic framework calls not for abandoning the earlier derivation but for enlarging the set of constraints. Adjoining to the continuous channel a jump channel that carries the arrival rate and the first two moments of the jump size, and re-running the same Maximum Entropy procedure, yields the Merton jump-diffusion process, with Geometric Brownian Motion recovered in the no-jump limit; the forward equation for the log-price density generalizes correspondingly from a Fokker--Planck to a Kolmogorov--Feller equation. On the pricing side the incompleteness of the market leaves the equivalent martingale measure no longer unique, and no-arbitrage alone cannot single one out; we show that imposing no-arbitrage through the mean log-return, the observable that already governs the dynamics, and maximizing the relative entropy of the pricing measure with respect to the jump-diffusion measure, selects the Esscher transform \cite{GerberShiu} from among the infinitely many equivalent martingale measures, here derived rather than borrowed. A different controlled observable would select a different measure, so that the non-uniqueness of the pricing measure becomes the non-uniqueness of the constraint through which no-arbitrage is imposed. The premium then obeys Merton's partial integro-differential equation, of which the Black--Scholes--Merton equation is the no-jump limit, and the resulting Poisson-weighted mixture of lognormals generates the implied-volatility smile and smirk from the very constraints that fix the dynamics.

The central contribution is methodological. Where most of the literature posits a stochastic model for the price, we derive the dynamics from the entropic formalism, and what distinguishes one model from another is never the inference but the information supplied to it. The contrast is that between Newton's theory of motion, in which the laws are postulated, and the Lagrangian formulation, in which they follow from a principle. The body of the paper develops this programme in two movements: the construction of the jump-diffusion dynamics together with its Kolmogorov--Feller equation, and then the pricing of European options, from the risk-neutral measure and Merton's equation to the volatility smile.

\section{Entropic Stock Model}

	The question we address is that of dynamics: given the present price of a stock, what can be said about how it will change? We do not inquire into how the market arrives at a price; we take the price as given and seek a principled account of its subsequent evolution.

\subsection{Return Symmetry}\label{subsec:return}
Before any constraint is imposed, we must identify the proper dynamical variable. The choice is not arbitrary; it is dictated by a symmetry of the problem itself.

In financial markets the relevant quantity is not the price of a security but its return. To an investor the absolute price is immaterial; what governs every allocation decision is the percentage gain or loss that follows from committing capital. A stock trading at a low price may well offer a superior return to one trading at a high price, and it is the return, not the price, that makes it the more attractive holding. The competitive mechanism of supply and demand enforces this principle: rising demand for the higher-returning stock bids up its price, eroding its return until, in equilibrium, securities of comparable risk offer comparable returns. The probability distribution that governs price evolution must therefore be invariant under any uniform rescaling of all prices; we call this requirement \emph{return symmetry}.

We now formalize this notion. Let $S>0$ denote the current price and $S^\prime > 0$ its subsequent value after an infinitesimal interval. For an arbitrary scale factor $\lambda > 0$, invariance of the transition probability under $S \mapsto \lambda S$ requires

\begin{equation}
P(\lambda S^\prime |\lambda S) \, d(\lambda S^\prime) = P(S^\prime | S) \, dS^\prime. \label{covlaw}
\end{equation}
Since $d(\lambda S^\prime) = \lambda \, dS^\prime$, this implies

\begin{equation}
P(\lambda S^\prime |\lambda S) = \frac{1}{\lambda} \, P(S^\prime | S), \hspace{5mm} \forall \lambda > 0. \label{covlaw2}
\end{equation}
Setting $\lambda = 1/S$ isolates the dependence on the return ratio $r := S^\prime / S$. With the return density $f(r) \equiv P(r | 1)$, normalized so that $\int_0^\infty f(r) \, dr = 1$, Equation~(\ref{covlaw2}) takes the canonical scale-invariant form

\begin{equation}
P(S^\prime |  S) = \frac{1}{S} \, f \! \left(  \frac{S^\prime}{S} \right). \label{returndensity}
\end{equation}
The content of (\ref{returndensity}) is that only the return $S^\prime / S$ determines the shape of the distribution; the prefactor $1/S$ is the Jacobian required by normalization and carries the sole dependence on the current price level.

Equation~(\ref{returndensity}) encodes the economic content of return symmetry, but it does not yet prescribe the most natural state variable for the analysis that follows. For the entropic construction of the prior, the directionality constraint, and the subsequent derivation of the Fokker--Planck equation, it is desirable to work in a variable that renders multiplicative rescalings additive. We now show that this requirement uniquely selects the logarithm of the price.

We seek a differentiable mapping $g: (0, \infty) \to \mathbb{R}$ such that the transition measure in the transformed variable is exactly invariant under $(S, S^\prime) \mapsto (\ell S, \ell S^\prime)$ for all $\ell > 0$:

\begin{equation}
P\big(g(S^\prime) | g(S)\big) \, d g(S^\prime) = P\big(g(\ell S^\prime) | g(\ell S)\big) \, d g(\ell S^\prime). \label{translationinv}
\end{equation}
Expanding $d g(\ell S^\prime) = g^\prime(\ell S^\prime) \, \ell \, d S^\prime$, Equation~(\ref{translationinv}) requires $\ell \, g^\prime(\ell S^\prime) = g^\prime(S^\prime)$ for all $\ell, S^\prime > 0$. Defining $h(S^\prime) := S^\prime \, g^\prime(S^\prime)$, this becomes $h(\ell S^\prime) = h(S^\prime)$, forcing $h$ to be constant. Denoting this constant by $C$, we arrive at the differential equation

\begin{equation}
S^\prime \, g^\prime(S^\prime) = C,
\end{equation}
whose general solution is $g(S) = C \ln S + D$. Setting $C=1$ and $D=0$ without loss of generality,

\begin{equation}
g(S) = \ln S.
\end{equation}
The log-price $\ln S$ is thus the unique differentiable variable (up to affine equivalence) for which the transition measure is exactly invariant under multiplicative rescaling. In this variable, the scale transformation $S \mapsto \ell S$ becomes the translation $\ln S \mapsto \ln S + \ln \ell$, and the transition density depends only on the log-price displacement,

\begin{equation}
P(\ln S^\prime | \ln S) = g\!\left( \ln \frac{S^\prime}{S} \right).
\end{equation}
Return symmetry has thereby selected both the dynamical variable and the structure of its transition density. The entropic construction that follows inherits this choice.

\subsection{The Continuous Channel}\label{subsec:continuous}

The dynamics of the log price is constructed in two stages, each governed by the same inferential principle. In this first stage we treat the price as evolving continuously; the result is a self-contained derivation of Geometric Brownian Motion that reproduces the continuous-only ED stock model of \cite{ABstock}. The second stage, which augments the microstate with a jump channel, is the subject of Section~\ref{subsec:jump}.

Our state of knowledge about the next log price is a probability distribution, and the relative entropy is the functional designed to update that distribution whenever new information becomes available. Given the current log price $\ln S$, we seek the transition density $P_c(\ln S^\prime | \ln S)$ over the interval $[t, t + \Delta t]$, where the subscript $c$ denotes the continuous channel throughout this section. We obtain $P_c$ by maximizing the relative entropy

\begin{equation}
\mathscr{S}[P_c,Q_c] = - \int \! d\ln S^\prime \;  P_c(\ln S^\prime \, | \, \ln S ) \, \ln \frac{P_c(\ln S^\prime \, | \, \ln S )}{Q_c(\ln S^\prime \, | \, \ln S )}, \label{relent}
\end{equation}
of the candidate density $P_c$ relative to a prior $Q_c$, subject to the constraints that encode the information at hand \cite{shore-johnson1980,Kevin}. We stress that the entropy referred to here is not that of physics; on the contrary, the thermodynamic entropy of statistical mechanics is itself derivable from inference \cite{Jaynes1, Jaynes2, Jaynes3}. When no substantive information is imposed beyond normalization, the posterior coincides with the prior, as it must, for in the absence of new information there is no ground on which to revise one's beliefs.

\subsubsection{The Prior}\label{sec:prior}
The prior $Q_c$ is itself the product of an inference: it is the distribution that already incorporates the information that motion is continuous, but is otherwise as uncommitted as possible; in particular, it carries no information about the direction of motion. Following the entropic-dynamics program \cite{ABstock}, we obtain $Q_c$ by maximizing its relative entropy relative to a reference $\Omega$ that encodes a state of complete ignorance about the next log price,

\begin{equation}
\mathscr{S}[Q_c, \Omega] = - \int \! d\ln S^\prime \; Q_c(\ln S^\prime \, | \, \ln S) \, \ln \frac{Q_c(\ln S^\prime \, | \, \ln S)}{\Omega(\ln S^\prime \, | \, \ln S)} . \label{rel_prior}
\end{equation}
Complete ignorance means the absence of any preference among regions of the state space; it is the assignment of equal probability to equal volumes of the log-price space. Since this space is the real line, which is flat, equal volumes are simply equal intervals of $\ln S^\prime$, and the reference reduces to a constant, $\Omega(\ln S^\prime \, | \, \ln S) = 1$. Any departure from uniformity would already embody information, which at this stage we do not possess. That $\Omega$ is not normalizable over the full real line is of no consequence: the continuity constraint confines the posterior to an infinitesimal neighbourhood of the current log price, so only the local behaviour of $\Omega$ matters, and within that neighbourhood a constant is as well-behaved as any density need be.

The first piece of information we incorporate is the \emph{continuity of motion}. It encapsulates the notion that any finite change in the log price is the accumulation of many infinitesimally small steps; the next log price is expected to remain close to its current value. In the language of inference, this is the information that is relevant to the continuous channel. A second channel, carrying the information that the log price may also undergo sudden discontinuous jumps, will be introduced in Section~\ref{subsec:jump}. We express continuity as a constraint on the mean-square log-price displacement,

\begin{equation}
\big\langle (\Delta \ln S)^2 \big\rangle_{Q_c} = \left\langle \left( \ln \frac{S^\prime}{S} \right)^2 \right\rangle_{Q_c} = k, \label{contconstraint}
\end{equation}
where $k$ is a small constant. In the limit $k \to 0$ the displacement vanishes and the next log price coincides with the current one, enforcing the continuity of motion. Maximizing (\ref{rel_prior}) subject to (\ref{contconstraint}) and normalization yields the driftless Gaussian

\begin{equation}
Q_c(\ln S^\prime \, | \, \ln S) = \sqrt{\frac{\alpha}{2\pi}} \, \exp\!\left[ -\tfrac{1}{2}\, \alpha \left( \ln \frac{S^\prime}{S} \right)^2 \right], \qquad \alpha = \frac{1}{k}, \label{prior_diff}
\end{equation}
whose spread is governed by the Lagrange multiplier $\alpha$ through $\big\langle (\ln(S^\prime/S))^2 \big\rangle_{Q_c} = 1/\alpha = k$. This is the same prior obtained in the continuous-only model \cite{ABstock}.

\subsubsection{The Entropic Clock}\label{sec:clock}
Inference by itself carries no notion of time. Before the framework can model dynamics, a concept of duration must be introduced from outside, and in Entropic Dynamics it enters as an \emph{entropic time} \cite{E-Time}, of which the time of physics is one particular realization. The Lagrange multiplier $\alpha$ conjugate to the continuity constraint provides precisely this: identifying the constraint value with the diffusion variance accumulated over the interval, $k = \sigma^2 \, \Delta t$ with $\sigma^2$ the volatility of the stock, turns $\alpha$ into the entropic clock

\begin{equation}
\alpha = \frac{1}{\sigma^2 \, \Delta t}, \label{diffclock}
\end{equation}
a notion of duration tailored to the system at hand. When the volatility is constant this clock coincides with Newtonian time \cite{NAC}; other identifications reproduce a relativistic notion of time \cite{IAC1, IAC2, IAC3}. In each case the entropic time is chosen to render the dynamics of the system as simple as possible.

\subsubsection{The Posterior and Geometric Brownian Motion}\label{sec:gbm}
The prior $Q_c$ is symmetric about the current log price: the continuity constraint tells us that the price will remain nearby, but says nothing about which direction it will tend to move. The second piece of information pertains to this \emph{directionality}. Stocks do not wander symmetrically; over any interval the log price exhibits a systematic tendency to drift, reflecting the expected performance of the underlying asset. We encode this information as a constraint on the expected log return and obtain the posterior $P_c(\ln S^\prime | \ln S)$ by maximizing the relative entropy (\ref{relent}) of $P_c$ relative to $Q_c$ subject to

\begin{equation}
\left< \ln \frac{S^\prime}{S} \right>_{P_c} = k_c,  \label{cons1}
\end{equation}
with $k_c$ to be determined. To arrive at $k_c$ we expand the log return to second order in $\Delta S / S$,

\begin{equation}
\ln \frac{S^\prime}{S} \approx \frac{\Delta S}{S} - \frac{1}{2}\left(\frac{\Delta S}{S}\right)^2, \label{Texp}
\end{equation}
and take the expectation with respect to the transition density. The first term is specified by the drift of the stock,

\begin{equation}
\left< \frac{\Delta S}{S} \right>_{P_c} = \mu_c \, \Delta t, \label{mu}
\end{equation}
where $\mu_c$ is the expected rate of return. The second term, to leading order, is fixed by the diffusion variance $k = \sigma^2 \Delta t$ already determined at the prior stage. Together they give $k_c = (\mu_c - \tfrac{1}{2}\sigma^2) \Delta t$ to leading order in $\Delta t$. Maximizing the relative entropy (\ref{relent}) of $P_c$ relative to the prior $Q_c$ of (\ref{prior_diff}), subject to the directionality constraint (\ref{cons1}) and normalization, yields the Gaussian transition density

\begin{equation}
P_c(\ln S^\prime \, | \, \ln S) = \frac{1}{\sqrt{2 \pi \sigma^2 \, \Delta t}} \, \exp \left[ - \frac{\big( \ln(S^\prime/S) - \mu_c \, \Delta t + \tfrac{1}{2} \sigma^2 \, \Delta t \big)^2}{2 \sigma^2 \, \Delta t} \right], \label{phi_c}
\end{equation}
the standard Geometric Brownian Motion transition density: the driftless prior (\ref{prior_diff}) whose mean has been shifted to $m_c \equiv (\mu_c - \tfrac{1}{2}\sigma^2) \Delta t$ by the directionality constraint.

The transition density (\ref{phi_c}) encodes the full probabilistic content of the continuous channel. We read off the dynamics by decomposing each log-price increment into its expected value and a zero-mean fluctuation,

\begin{equation}
\Delta \ln S = \langle \Delta \ln S \rangle_{P_c} + \Delta W, \label{drift_fluct}
\end{equation}
where the drift is read off from (\ref{phi_c}),

\begin{equation}
\langle \Delta \ln S \rangle_{P_c} = (\mu_c - \tfrac{1}{2}\sigma^2)\,\Delta t, \label{cont_drift}
\end{equation}
and the fluctuation $\Delta W$ satisfies

\begin{equation}
\langle \Delta W \rangle_{P_c} = 0, \qquad \langle \Delta W^2 \rangle_{P_c} = \frac{1}{\alpha} = \sigma^2 \, \Delta t. \label{fluct}
\end{equation}
The entropic clock (\ref{diffclock}) is what makes the fluctuation variance scale linearly with the interval; over a short step the fluctuation $\Delta W \sim \mathcal{O}(\sigma \sqrt{\Delta t})$ dominates the drift $\mathcal{O}(\Delta t)$, producing a motion that is continuous but nowhere differentiable. In the continuum limit the log price obeys

\begin{equation}
\Delta \ln S = (\mu_c - \tfrac{1}{2} \sigma^2) \, \Delta t + \Delta W, \label{gbm_logSDE}
\end{equation}
Written in terms of the price itself, the same dynamics reads

\begin{equation}
\Delta S = \mu_c \, S \, \Delta t + S \, \Delta W, \label{gbm_SDE}
\end{equation}
the Geometric Brownian Motion \cite{ABstock}. It is this process that Black, Scholes, and Merton \cite{BS, Merton} adopted as their starting point, postulating it as an ansatz on which the celebrated Black--Scholes option-pricing framework is built. Here the same process has been \emph{derived}, not assumed, from two constraints alone, continuity and directionality, imposed on a single microstate through the relative entropy.

The continuous channel, however, confines all price movements to the Gaussian regime and cannot account for the sudden discontinuous moves observed in real markets. To accommodate such events we must enlarge the microstate and introduce a second channel.

\subsection{The Jump Channel}\label{subsec:jump}

The Geometric Brownian Motion derived above captures the small, continuous fluctuations that dominate routine trading, but it cannot account for the sudden, large moves triggered by discrete information events: earnings surprises, central-bank decisions, geopolitical shocks. Empirically, asset returns are leptokurtic and negatively skewed \cite{Fama, Cont01stylized}, and the implied-volatility surface exhibits a persistent smile and smirk \cite{Bates96}. These departures from the Gaussian are not artifacts; they are the signature of jumps. A virtue of the entropic framework is that it is modular in its constraints: each piece of information enters through a well-defined constraint, and the formalism itself remains unchanged. When new information becomes available, one does not discard the existing model and start afresh; one enlarges the microstate to accommodate the new degrees of freedom, imposes the additional constraints that encode the new information, and updates the distribution by maximizing the relative entropy exactly as before. The inference engine is always the same; what changes from one model to the next is solely the information that is taken into account. Extending the continuous channel to include jumps is therefore not a departure from the framework but a natural application of it.

\subsubsection{The Augmented Microstate and Jump Constraints}\label{sec:augmented}
In the continuous channel the microstate was the next log price $\ln S^\prime$ alone. To incorporate jumps we enlarge the microstate to the pair $(\ln S^\prime, n)$, where $n$ is the number of jump arrivals during $\Delta t$. In principle $n$ may take any non-negative integer value, but in the regime of short intervals that defines the entropic instant, at most one discrete event can occur before the next update; simultaneous arrivals are excluded by the same logic that restricts the continuous channel to infinitesimally small steps. The short-step count is consequently \emph{binary}, $n \in \{0, 1\}$: either no jump occurs and the price evolves by diffusion alone, which recovers the continuous-channel microstate of the first stage, or exactly one jump is superposed on the diffusive motion. This Bernoulli structure is not an approximation imposed for convenience; it is the exact leading-order content of the arrival process. Over a finite horizon the independent short-step counts accumulate, and their sum follows the Poisson distribution derived in \ref{app:poisson_finite}.

The prior for this second stage is dictated by the sequential logic of entropic inference, in which the posterior of one step serves as the prior of the next. The continuous posterior $P_c$ of (\ref{phi_c}), already obtained, becomes the reference on the continuous channel; the new count variable, about which nothing is yet known, carries a uniform reference $\omega_n$ on $\{0,1\}$. The augmented prior is therefore $P_c(\ln S^\prime | \ln S)\,\omega_n$.

Three constraints, absent from the first stage, now encode the jump information. The resulting joint posterior over the augmented microstate $(\ln S^\prime, n)$ will be denoted $P_{JD}$, where the subscript marks the jump-diffusion channel to distinguish it from the continuous posterior $P_c$. The expected number of arrivals per interval fixes the jump intensity,

\begin{equation}
\left< n \right>_{P_{JD}} = \lambda \, \Delta t, \label{lambdacons}
\end{equation}
with $\lambda > 0$. When a jump occurs, the log return decomposes into its continuous and discontinuous parts,

\begin{equation}
\Delta \ln S = z + Y, \label{jumpdecomp}
\end{equation}
where $z$ is the continuous displacement distributed by $P_c$ and $Y$ is the jump increment in the log price. Because the microstate selected by scale invariance is $\ln S$, the jump size $Y$ is itself a log-price displacement; it measures the ratio of the post-jump price to the pre-jump price on a logarithmic scale. Our information about the magnitude of jumps is encoded through the expected value and the expected square of $Y$,

\begin{equation}
\left< Y \right>_{P_{JD}} = \kappa, \qquad \left< Y^2 \right>_{P_{JD}} = \nu, \label{jumpconstraints}
\end{equation}
with $\kappa$ the mean log-price jump and $\nu$ its second moment. The continuous drift has already been determined by the directionality constraint (\ref{cons1}); the new constraints (\ref{lambdacons}) and (\ref{jumpconstraints}) bear exclusively on the jump count and the jump size. It is this separation, the absence of any constraint coupling the jump to the diffusion, that will produce the factorization of the posterior.

\subsubsection{The Merton Jump-Diffusion Transition Density}\label{sec:constraints}
One could impose all three jump constraints simultaneously and arrive at the same posterior $P_{JD}$; the final result does not depend on the order of updating. We choose instead to impose the constraints in sequence, so that the inferential role of each is transparent. Both updates maximize the relative entropy of the joint distribution relative to the running prior,

\begin{equation}
\mathscr{S}[P_{JD}, P_c\omega_n] = - \sum_{n} \int \! d\ln S^\prime \, P_{JD}(\ln S^\prime, n | \ln S) \, \ln \frac{P_{JD}(\ln S^\prime, n | \ln S)}{P_c(\ln S^\prime | \ln S)\,\omega_n}, \label{relent_joint}
\end{equation}
the functional (\ref{relent}) extended by the sum over the count.

\subsubsection*{The Arrival Constraint}
We begin with the arrival constraint (\ref{lambdacons}) alone, $\langle n \rangle_{P_{JD}} = \lambda \, \Delta t$, leaving the jump sizes unconstrained. Maximizing the relative entropy (\ref{relent_joint}) subject to this single constraint and normalization returns the Gibbs form $P_{JD} \propto P_c\,\omega_n\,e^{-\gamma n}$, in which $\gamma$ is the Lagrange multiplier conjugate to the arrival constraint. Since the tilt $e^{-\gamma n}$ depends on the count alone, it leaves the log-price factor untouched, and the joint distribution factorizes exactly,

\begin{equation}
P_{JD}(\ln S^\prime, n \, | \, \ln S) = P(n) \; P_c(\ln S^\prime | \ln S), \label{count_only}
\end{equation}
with $P(n) \propto \omega_n\,e^{-\gamma n}$. The count therefore lives on the binary space $\{0,1\}$ of the entropic instant, where normalization gives $P(0) = 1/(1 + e^{-\gamma})$ and $P(1) = e^{-\gamma}/(1 + e^{-\gamma})$. The multiplier is then determined by the constraint itself: imposing $\langle n \rangle = P(1) = \lambda \, \Delta t$ and solving for $\gamma$,

\begin{equation}
e^{-\gamma} = \frac{\lambda \, \Delta t}{1 - \lambda \, \Delta t}, \qquad \gamma = \ln \frac{1 - \lambda \, \Delta t}{\lambda \, \Delta t} = \ln \frac{P(0)}{P(1)}, \label{gammasol}
\end{equation}
so that $\gamma$ is fixed in closed form by the arrival rate $\lambda$ and the interval $\Delta t$. The multiplier carries a transparent meaning: it is the log-odds against a jump over the interval, large and positive when arrivals are rare, vanishing when a jump is as likely as not. In this it is the discrete-count counterpart of the diffusion multiplier $\alpha = 1/(\sigma^2 \Delta t)$, which sets the scale of the continuous channel. Substituting (\ref{gammasol}) back into the normalized probabilities collapses them to the Bernoulli short-step count

\begin{equation}
P(0) = 1 - \lambda \, \Delta t, \qquad P(1) = \lambda \, \Delta t. \label{bernoullicount}
\end{equation}
Over a finite horizon the independent short steps accumulate, and their sum follows the Poisson distribution

\begin{equation}
P(n) = \frac{(\lambda \, \Delta t)^n}{n!} \, e^{-\lambda \, \Delta t}, \label{poisson}
\end{equation}
obtained from the arrival constraint alone in \ref{app:poisson_finite}.

One consequence of the factorization (\ref{count_only}) merits emphasis. Because the joint posterior is the bare product $P(n)\,P_c$, the marginal distribution of the log price, recovered by summing over the count, is nothing but $P_c$ itself: the Geometric Brownian Motion density of the continuous channel, entirely unaltered. The arrival constraint has supplied a new degree of freedom, the count $n$, and settled its distribution, yet it has said nothing about what a jump does to the price. Until the magnitude of the jumps is specified, the price distribution cannot tell a world with jumps from one without. Only with the jump-size constraints, imposed in the next update, does the price distribution part ways with Geometric Brownian Motion. Setting $\lambda = 0$ retires the count variable altogether and restores the first-stage dynamics exactly.

\subsubsection*{The Jump-Size Constraints}
With the count distribution settled, we promote the count-only posterior (\ref{count_only}) to the prior of the next update and bring in the remaining information: the jump-size constraints (\ref{jumpconstraints}) on the first two moments of $Y$. On the one-jump branch the log return resolves through (\ref{jumpdecomp}) into the pair $(z, Y)$ of a continuous displacement and a jump, and the running prior on this pair is $P(n)\,P_c(z)\,\omega_Y$, with $\omega_Y = 1$ the uniform reference on the jump size, as yet unconstrained. Since the constraints (\ref{jumpconstraints}) involve $Y$ and nothing else, maximizing the relative entropy subject to them and normalization reweights the jump variable alone, tilting it exponentially while leaving the count and the diffusion untouched,

\begin{equation}
P_{JD}(z, Y, n \, | \, \ln S) = \frac{1}{Z}\, P(n)\, P_c(z)\, \omega_Y\, e^{-b Y - c Y^2}, \label{stationarity_joint}
\end{equation}
where $b$ and $c$ are the multipliers conjugate to the mean and the second moment of $Y$. The count $P(n)$ and the continuous factor $P_c(z)$ are carried through verbatim, and the no-jump branch, bearing no size constraint, coincides with its prior $P(0)\,P_c$. The posterior therefore separates cleanly into three factors, one for each channel of information, the count, the continuous displacement, and the jump,

\begin{equation}
P_{JD}(\ln S^\prime, n \, | \, \ln S) = P(n) \; P_{JD}(\ln S^\prime \, | \, \ln S, \, n), \label{joint_factored}
\end{equation}
in which $P(n)$ is the Bernoulli count (\ref{bernoullicount}) over the short step, its Poisson accumulation (\ref{poisson}) over a finite horizon, and $P_{JD}(\ln S^\prime | \ln S, n)$ the branch-conditional return density obtained below. We stress that this separation is \emph{derived, not posited}: it is forced by the structure of the constraints, none of which couples the jump to the diffusion. Independence between the two channels is thus a theorem of the joint Maximum Entropy procedure rather than an assumption laid down in advance.

\subsubsection*{The Single-Jump Density}
The single-jump distribution is now read directly from the tilted posterior (\ref{stationarity_joint}). Isolating the factor that carries the jump variable gives $P_J(Y) = Z_J^{-1}\,\omega_Y\,e^{-bY - cY^2}$, a quadratic exponential that, on completing the square, is at once recognized as a Gaussian of mean $-b/(2c)$ and variance $1/(2c)$. The two multipliers are then eliminated in favour of the quantities they were introduced to enforce: the moment constraints (\ref{jumpconstraints}) fix the mean at $\kappa$ and the variance at $\delta^2 \equiv \nu - \kappa^2$, so that

\begin{equation}
P_J(Y) = \frac{1}{\sqrt{2 \pi \delta^2}} \, \exp \left[ -\frac{(Y - \kappa)^2}{2 \delta^2} \right]. \label{mertonjump}
\end{equation}

This is the Gaussian jump-size distribution of Merton's jump-diffusion model \cite{Merton76jump}. Here it is not assumed but obtained, as the unique distribution consistent with the two moment constraints on the jump size and otherwise maximally noncommittal; the multipliers $b$ and $c$ are precisely what translate those constraints into the mean $\kappa$ and the dispersion $\delta$ of the jump.

\subsubsection*{The Merton Jump-Diffusion Density}
It remains to assemble the branch-conditional return density from its two independent constituents. Conditioned on exactly $n$ arrivals, the log return (\ref{jumpdecomp}) is the sum $\ln(S^\prime/S) = z + \sum_{i=1}^{n} Y_i$ of a continuous displacement $z$ and $n$ jumps, each governed by the Gaussian (\ref{mertonjump}). These contributions are mutually independent, not by fiat but as a consequence of the inference itself: the factorization (\ref{joint_factored}), in which no constraint couples the jump to the diffusion, renders $z$ and $Y$ independent, while the independent accumulation of the entropic short steps established in \ref{app:poisson_finite} renders the successive jumps independent of one another. The density of a sum of independent quantities is the convolution of their densities; the conditional return density is therefore the diffusion kernel convolved with $n$ copies of the single-jump distribution. Since every constituent is Gaussian, the convolution is again Gaussian, its means and variances simply adding,

\begin{align}
P_{JD}(\ln S^\prime \, | \, \ln S, n) &= \big[\, P_c \, * \, P_J^{*n} \,\big]\!(\ln S^\prime - \ln S) \nonumber \\
&= \frac{1}{\sqrt{2\pi(\sigma^2 \Delta t + n \delta^2)}} \, \exp\!\left[ -\frac{\big(\ln(S^\prime/S) - m_c - n \kappa\big)^2}{2(\sigma^2 \Delta t + n \delta^2)} \right], \label{conditional_n}
\end{align}
where $m_c = (\mu_c - \tfrac{1}{2}\sigma^2)\Delta t$ is the continuous drift, $*$ denotes convolution in the log-price displacement, and $P_J^{*n}$ is the $n$-fold self-convolution of the single-jump density. For $n = 0$ this self-convolution is the identity element of the convolution, the Dirac measure, which reproduces its argument unchanged; the no-jump branch therefore returns the bare diffusion kernel $P_c$. Each additional jump shifts the conditional mean by $\kappa$ and broadens the conditional variance by $\delta^2$.

Marginalizing this conditional density over the count $n$, with weights $P(n)$, restores the full transition density. Over a single short step the binary count (\ref{bernoullicount}) admits at most one arrival, and the average collapses to two terms,

\begin{equation}
P_{JD}(\ln S^\prime \, | \, \ln S) = (1 - \lambda \, \Delta t) \, P_c(\ln S^\prime - \ln S) + \lambda \, \Delta t \, \big[\, P_c \, * \, P_J \,\big]\!(\ln S^\prime - \ln S),  \label{infinitesimal}
\end{equation}
which exhibits the two-channel architecture in its starkest form: with probability $1 - \lambda \, \Delta t$ the price advances by diffusion alone, while with probability $\lambda \, \Delta t$ a single jump is superposed upon that diffusion. The representation is exact at the binary level, free of any $\mathcal{O}((\Delta t)^2)$ remainder. Over a finite horizon the accumulated Poisson count (\ref{poisson}) replaces the two-term average by the series

\begin{equation}
P_{JD}(\ln S^\prime \, | \, \ln S) = \sum_{n=0}^{\infty} P(n) \, \big[\, P_c \, * \, P_J^{*n} \,\big]\!(\ln S^\prime - \ln S), \label{marginalized}
\end{equation}
with $\Delta t$ replaced by the maturity $T$ when the parameters are held uniform. This Poisson-weighted superposition of Gaussians is precisely the Merton jump-diffusion transition density \cite{Merton76jump}, here derived rather than postulated. Setting $\lambda = 0$ annihilates every term beyond $n = 0$ and collapses both (\ref{infinitesimal}) and (\ref{marginalized}) to the Geometric Brownian Motion density (\ref{phi_c}), recovering the continuous-only model exactly.

\subsubsection{The Merton Jump-Diffusion Process}\label{sec:sde}
The stochastic process underlying the transition density is recovered by the same drift-plus-fluctuation decomposition that delivered Geometric Brownian Motion in Section~\ref{sec:gbm}. On the continuous branch the density is the bare kernel $P_c$, so the decomposition (\ref{drift_fluct}) carries over verbatim. The jump channel enters through the additive structure of the log return (\ref{jumpdecomp}): over an infinitesimal step the count contributes a single displacement $Y \sim P_J$ with probability $\lambda \, \Delta t$, and the log price advances by the compound increment

\begin{equation}
\Delta \ln S = (\mu_c - \tfrac{1}{2} \sigma^2) \, \Delta t + \Delta W + Y \, \Delta N, \label{jd_logSDE}
\end{equation}
in which $W$ is the Wiener process of (\ref{drift_fluct}), with $\langle \Delta W^2 \rangle = \sigma^2 \, \Delta t$, and $N$ is the counting process with $\langle \Delta N \rangle = \lambda \, \Delta t$. Equation (\ref{jd_logSDE}) is a short-step relation, written over an infinitesimal interval $\Delta t$: there the increment $\Delta N \in \{0, 1\}$ is Bernoulli, equal to one with probability $\lambda \, \Delta t$ and zero otherwise, the chance of two or more arrivals being $\mathcal{O}((\Delta t)^2)$ and neglected, so that at most one displacement $Y \sim P_J$ is appended per step. Accumulated over a finite horizon these Bernoulli steps build the Poisson count $N_t$ (\ref{app:poisson_finite}), and the jumps $J_t \equiv \sum_{k=1}^{N_t} Y_k$ constitute a compound-Poisson process. Expressed in terms of the price itself, the same dynamics reads

\begin{equation}
\frac{\Delta S}{S} = \mu_c \, \Delta t + \Delta W + (e^{Y} - 1) \, \Delta N, \label{jd_SDE}
\end{equation}
In this passage from the log price to the price, the additive log-jump $Y$ becomes the proportional price jump $e^{Y} - 1$, and the $-\tfrac{1}{2}\sigma^2$ term is reabsorbed into the drift. This is the \emph{Merton jump-diffusion} \cite{Merton76jump}, obtained here not by postulate but by derivation: it is the process that emerges when five pieces of information, two governing the continuous channel and three the jump channel, are imposed through the relative entropy.

The two single-channel limits stand at either end of this construction. Setting $\lambda = 0$ silences the jump term and restores the Geometric Brownian Motion (\ref{gbm_logSDE})--(\ref{gbm_SDE}); setting $\sigma = 0$ silences the diffusion and leaves a pure compound-Poisson process. The Merton jump-diffusion therefore unites both channels in a single dynamics, with Geometric Brownian Motion and the pure-jump process recovered as its two limiting cases.

\subsection{Entropic Instant and Probability Dynamics}\label{subsec:KF}
To complete the notion of entropic time we must say what is meant by an instant. In Entropic Dynamics an instant is nothing other than a state of knowledge, a probability distribution, and time advances through the construction of each new instant from the one preceding it. The \emph{entropic instant} that effects this construction is defined by

\begin{equation}
p(\ln S^\prime)_{t^\prime} = \int d \ln S \, P_{JD}(\ln S^\prime \, | \, \ln S) \, p(\ln S)_t,  \label{CK}
\end{equation}
in which $P_{JD}(\ln S^\prime \, | \, \ln S)$ is the two-channel short-step transition density (\ref{infinitesimal}) furnished by the joint Maximum Entropy procedure of Section~\ref{sec:constraints}. The distribution $p(\ln S)_t$ represents the whole of the information available at the instant $t$, and the next instant is the distribution $p(\ln S^\prime)_{t^\prime}$ generated from it; for brevity we henceforth write $p(\ln S, t)$ in place of $p(\ln S)_t$. It bears noting that the count $n$ enters neither (\ref{CK}) nor the forward equation to which it gives rise. The kernel (\ref{infinitesimal}) is already marginal over the count, the jump arrivals having been summed out in the passage from (\ref{joint_factored}) to (\ref{infinitesimal}); the propagated object is therefore the marginal density of the log price alone, bearing no count index, and the jumps persist only through the integral operator that the marginalization bequeaths to the dynamics. The time parameter so constructed is ordered and carries an arrow \cite{E-Time}. Our task in this subsection is to extract from (\ref{CK}) the forward equation governing the evolution of $p(\ln S, t)$. That equation is the \emph{Kolmogorov--Feller equation}, the forward equation proper to the two-channel process of Section~\ref{sec:sde}, of which the \emph{Fokker--Planck equation} of the original continuous-only stock model is the no-jump limit.

The differential form of the entropic-instant identity (\ref{CK}) is reached by substituting the infinitesimal two-channel kernel (\ref{infinitesimal}), expanding the diffusion convolution to first order in $\Delta t$, and letting $\Delta t \to 0$. The result is the \emph{Kolmogorov--Feller equation}, written here with $y$ denoting a value of the log-price jump $Y$ of (\ref{mertonjump}) and the integration taken against the jump-size density $P_J$,

\begin{align}
\partial_t p(\ln S, t) = & -\frac{\partial}{\partial \ln S} \!\left[ \! \big( \mu_c - \tfrac{1}{2} \sigma^2 \big) \, p(\ln S, t) \! \right] + \frac{1}{2} \frac{\partial^2}{\partial (\ln S)^2} \!\left[ \sigma^2 \, p(\ln S, t) \right] \nonumber \\
& + \lambda \!\int \! dy \, \big[ p(\ln S - y, t) - p(\ln S, t) \big] \, P_J(y),  \label{KF}
\end{align}
the forward equation for the probability density on the microstate; its step-by-step derivation from the entropic-instant identity (\ref{CK}) is given in \ref{app:KF}. The right-hand side resolves into two operators acting on $p$. The first, the \emph{diffusion operator} formed by the two derivative terms, is the familiar Fokker--Planck differential operator; the second, the \emph{jump operator} given by the integral, registers the influx of probability into $\ln S$ from the sources $\ln S - y$, balanced against the outflux from $\ln S$. The two operators are the imprint of the two channels of the microstate, the diffusion operator carrying the continuous channel and the jump operator the discontinuous one. With $P_J$ the Gaussian (\ref{mertonjump}), equation (\ref{KF}) is the forward equation of Merton's jump-diffusion process \cite{Merton76jump}.

Setting $\lambda = 0$ in (\ref{KF}) silences the integral term and leaves precisely the diffusion operator already displayed there, namely the Fokker--Planck equation for the time-evolved log-price density of the original continuous-only stock model; the Fokker--Planck equation governing Geometric Brownian Motion is thus recovered as the no-jump limit of the present derivation. Conversely, letting $\sigma^2 \to 0$ in (\ref{KF}) silences the diffusion operator and leaves the master equation of a pure compound-Poisson process on the log price. The Kolmogorov--Feller equation therefore unites the two channels at the level of the density, with the Fokker--Planck equation and the pure-jump master equation recovered as its two limiting cases.


\section{European Option Pricing}

Having derived the dynamics of the stock, we now turn to the valuation of European options written upon it. The guiding principle is unchanged: options are priced by the same Maximum Entropy method that produced the dynamics, the sole new ingredient being the information that the market admits no arbitrage. Since the jump market is incomplete its equivalent martingale measures are infinitely many, and no-arbitrage alone selects none; imposing it through the mean log-return and maximizing the relative entropy from the jump-diffusion measure selects the Esscher family, whose unique member is fixed by the martingale condition (Section~\ref{subsec:Esscher}). The exponential tilt so obtained is the Esscher transform, here the output of the inference rather than imposed from without. The classical Black--Scholes risk-neutral measure, together with the Black--Scholes formulas for the call and the put, is recovered as the no-jump limit $\lambda = 0$ of this construction. The premium itself is governed by Merton's partial integro-differential equation (PIDE), which we derive in Section~\ref{subsec:MertonPIDE} from the backward Kolmogorov--Feller equation under $P_{\mathrm{rn}}$; the Black--Scholes--Merton differential equation is in turn its $\lambda_{\mathrm{rn}} = 0$ limit. As with the dynamics, what distinguishes one pricing model from another is not the method but the information supplied to it.


\subsection{Risk-Neutral Valuation}\label{subsec:Esscher}
A derivative security must be priced so as to leave no arbitrage opportunity. \emph{Arbitrage} is the possibility of a self-financing trading strategy that costs nothing to set up, can never lose money, and yields a strictly positive profit with positive probability: a riskless ``free lunch.'' The no-arbitrage principle is the assumption that such opportunities do not persist in a competitive market, because any that appeared would be exploited and competed away almost at once. It is the weakest and most widely accepted postulate in the theory of asset pricing: it presupposes only that market participants prefer more wealth to less, and not any particular model of preferences, utility, or market equilibrium.

It is worth emphasizing that no-arbitrage forbids only \emph{riskless} profit; it does not forbid risk-taking. An investor is free to hold a risky position and earn, or lose, a return commensurate with that risk. What is excluded is a position that bears \emph{no} risk yet earns more than the risk-free rate $r_f$, or that earns a strictly positive payoff for zero net outlay. Ordinary speculation, whose gains are uncertain, is entirely consistent with no-arbitrage; only the \emph{certain}, costless gain is ruled out. ``No arbitrage'' therefore means no riskless arbitrage, and risky bets remain perfectly admissible.

The bridge from this economic principle to a probability measure on the microstate is the Fundamental Theorem of Asset Pricing \cite{HarrisonKreps, DelbaenSchachermayer}: a market is free of arbitrage if and only if there exists an \emph{equivalent martingale measure}, a measure $P_{\mathrm{rn}}$ that assigns probability zero to exactly the same events as the jump-diffusion measure $P_{JD}$ and under which the discounted price process is a martingale. In one-step form this is the single scalar condition
\begin{equation}
\left\langle \frac{S^\prime}{S} \right\rangle_{\mathrm{rn}} = \big\langle e^{x} \big\rangle_{\mathrm{rn}} = e^{r_f \, \Delta t}, \qquad x \equiv \ln \frac{S^\prime}{S}. \label{martingale}
\end{equation}
When $\lambda = 0$ the market is complete and this measure is unique; once $\lambda > 0$ the jump risk borne by the compound-Poisson channel can no longer be hedged away by trading the underlying alone, the market is incomplete, and the equivalent martingale measures form an infinite family. Condition (\ref{martingale}) is then one scalar restriction on an infinite-dimensional set of admissible measures: it constrains a pricing measure but cannot single one out, and a further principle is required. The entropic framework supplies it by the same rule of inference used throughout this work, the maximization of relative entropy from $P_{JD}$; the only question, taken up next, is the observable through which no-arbitrage is to be imposed.

\subsubsection*{No-Arbitrage as a Constraint on the Mean Log-Return}

No-arbitrage is most naturally a statement about the arithmetic return. Since $\langle S^\prime/S \rangle_{\mathrm{rn}} = e^{r_f \Delta t}$,
\begin{equation}
\Big\langle \frac{\Delta S}{S} \Big\rangle_{\mathrm{rn}} = \Big\langle \frac{S^\prime}{S} \Big\rangle_{\mathrm{rn}} - 1 = e^{r_f \Delta t} - 1, \label{noarb_arith}
\end{equation}
the exact counterpart of the drift constraint (\ref{mu}), with the growth rate $\mu_c$ replaced by the risk-free rate $r_f$. The observable the entropic update will control is the log return, and we pass to it by the same expansion that carried the forward derivation, (\ref{Texp}), now taken under the pricing measure,
\begin{equation}
\Big\langle \ln \frac{S^\prime}{S} \Big\rangle_{\mathrm{rn}} = \Big\langle \frac{\Delta S}{S} \Big\rangle_{\mathrm{rn}} - \frac{1}{2} \Big\langle \Big( \frac{\Delta S}{S} \Big)^{\!2} \Big\rangle_{\mathrm{rn}} + \cdots . \label{jensen}
\end{equation}
In the diffusion limit $\lambda = 0$ the displacement is of order $\sqrt{\Delta t}$, the expansion terminates at second order with $\langle (\Delta S/S)^2 \rangle_{\mathrm{rn}} = \sigma^2 \Delta t$, and (\ref{noarb_arith})--(\ref{jensen}) return the risk-neutral log-drift
\begin{equation}
\Big\langle \ln \frac{S^\prime}{S} \Big\rangle_{\mathrm{rn}} = \big( r_f - \tfrac{1}{2}\sigma^2 \big)\Delta t, \label{jensen_diff}
\end{equation}
exactly the directionality result of Section~\ref{sec:gbm}, the constraint (\ref{cons1}) run with $\mu_c \to r_f$. Once $\lambda > 0$ the jump displacement $\Delta S/S = e^{Y} - 1$ is of order unity, the second-order truncation no longer suffices, and the expansion must be carried to all orders. Retaining it, the mean log-return reads
\begin{equation}
\Big\langle \ln \frac{S^\prime}{S} \Big\rangle_{\mathrm{rn}} = \big( r_f - \tfrac{1}{2}\sigma^2 \big)\Delta t \;-\; \lambda_{\mathrm{rn}} \big\langle e^{Y} - 1 - Y \big\rangle_{\mathrm{rn}} \, \Delta t, \label{jensen_resolved}
\end{equation}
the jump term being the convexity correction the discrete channel contributes; it vanishes with $\lambda$ and restores (\ref{jensen_diff}) in the diffusion limit. Here $\lambda_{\mathrm{rn}}$ and $\langle\,\cdot\,\rangle_{\mathrm{rn}}$ are the yet-to-be-determined risk-neutral quantities the constraint must satisfy self-consistently; that it can be written in this two-channel form already anticipates the structure-preserving measure constructed below, the general constraint being (\ref{jensen}). Like the drift constraint (\ref{mu}) from which it descends, (\ref{jensen_resolved}) holds to first order in $\Delta t$, terms of $\mathcal{O}((\Delta t)^2)$ being neglected. The point to retain is that (\ref{jensen}) constrains the mean log-return, the very observable whose mean the directionality constraint (\ref{cons1}) fixed when the dynamics were derived. Imposing no-arbitrage therefore revises that single number, replacing the subjective growth rate by the risk-free rate adjusted for convexity, and calls on no dynamical observable beyond those already present in the model.

\subsubsection*{The Maximum-Entropy Update and the Esscher Transform}

We update the jump-diffusion measure by maximizing its relative entropy
\begin{equation}
\mathscr{S}[P_{\mathrm{rn}}, P_{JD}] = - \int \! d\ln S^\prime \; P_{\mathrm{rn}}(\ln S^\prime \,|\, \ln S) \, \ln \frac{P_{\mathrm{rn}}(\ln S^\prime \,|\, \ln S)}{P_{JD}(\ln S^\prime \,|\, \ln S)}, \label{relent_rn}
\end{equation}
subject to normalization and to the mean-log-return constraint (\ref{jensen}). The prior is the marginalized jump-diffusion density of (\ref{marginalized}), with the count $n$ already summed out, since no-arbitrage constrains the price alone and hence only the marginal distribution of $x$. Introducing a single Lagrange multiplier $\theta$ conjugate to $\langle x \rangle$, maximizing the relative entropy yields the Gibbs form, an exponential tilt of the prior along the log return,
\begin{equation}
P_{\mathrm{rn}}(\ln S^\prime \,|\, \ln S) = \frac{1}{Z}\, e^{\theta x}\, P_{JD}(\ln S^\prime \,|\, \ln S), \qquad Z = Z(\theta) = \big\langle e^{\theta x} \big\rangle_{P_{JD}}. \label{esscher_tilt}
\end{equation}
This is exactly the transformation the actuarial and L\'evy-process literature calls the \emph{Esscher transform} \cite{GerberShiu}. The order of the argument deserves emphasis: the Esscher transform is not postulated and afterwards justified, nor is it asserted to follow from no-arbitrage alone; it is the Maximum Entropy update of the jump-diffusion measure under the mean-log-return constraint into which no-arbitrage was recast, and the name attaches to it only in hindsight. That the controlled observable is the log return, and not the gross return, is what makes the tilt exponential in $x$ and hence structure-preserving, as the channel decomposition now shows.

The marginalized density (\ref{marginalized}) is a Poisson-weighted mixture of Gaussians, and on the branch with $n$ jumps the log return is the sum $x = z + \sum_i y_i$ of a continuous displacement and the jumps, so the tilt factorizes, $e^{\theta x} = e^{\theta z}\prod_i e^{\theta y_i}$, and reweights each constituent on its own. The continuous channel is tilted by $e^{\theta z}$, which shifts the mean of the Gaussian displacement and leaves its variance untouched, so the continuous drift moves to
\begin{equation}
\mu_{c,\mathrm{rn}} = \mu_c + \theta \sigma^2, \label{rn_drift_tilt}
\end{equation}
the volatility $\sigma$ preserved. The jump-size density is reweighted and renormalized,
\begin{equation}
P_{J,\mathrm{rn}}(y) = \frac{e^{\theta y} \, P_J(y)}{\int e^{\theta y^\prime} P_J(y^\prime) \, dy^\prime},  \label{esscher_density}
\end{equation}
and, since each jump contributes a factor $M_J(\theta) \equiv \int e^{\theta y} P_J(y)\, dy$, the tilt multiplies the $n$-jump branch of the mixture by $M_J(\theta)^n$ and rescales the Poisson weights, sending the arrival intensity to
\begin{equation}
\lambda_{\mathrm{rn}} = \lambda \int e^{\theta y} \, P_J(y) \, dy = \lambda\, M_J(\theta). \label{esscher_intensity}
\end{equation}
These are not three new postulates but the single tilt $e^{\theta x}$ resolved into its channels: under $P_{\mathrm{rn}}$ the process is once more a jump-diffusion, of the very form (\ref{infinitesimal}) with $(\mu_c, \lambda, P_J)$ replaced by $(\mu_{c,\mathrm{rn}}, \lambda_{\mathrm{rn}}, P_{J,\mathrm{rn}})$ and the volatility preserved. This structure preservation is the property the log-return constraint secures and a constraint on the gross return would forfeit.

\subsubsection*{Fixing the Multiplier}

Only the multiplier $\theta$ remains free, and it is fixed by requiring the constraint (\ref{jensen_resolved}) to hold self-consistently, its right-hand side being evaluated under $P_{\mathrm{rn}}$ itself. Equivalently, since $S^\prime/S = e^{x}$, the martingale expectation (\ref{martingale}) is, under the tilt (\ref{esscher_tilt}), the ratio of normalizers $Z(\theta+1)/Z(\theta)$. The normalizer factorizes over the channels,
\begin{align}
Z(\theta) &= M_c(\theta)\, \exp\!\big[ \lambda \Delta t\,(M_J(\theta) - 1) \big], \label{rn_Z} \\
M_c(\theta) &= \exp\!\big[ \theta(\mu_c - \tfrac{1}{2}\sigma^2)\Delta t + \tfrac{1}{2}\theta^2 \sigma^2 \Delta t \big], \label{rn_Mc}
\end{align}
with $M_c$ the moment generating function of the continuous Gaussian channel and $M_J$ that of the jump size, so the ratio evaluates to
\begin{equation}
\Big\langle \tfrac{S^\prime}{S} \Big\rangle_{\mathrm{rn}} = \frac{Z(\theta+1)}{Z(\theta)} = \exp\!\Big[ \big( \mu_c + \theta \sigma^2 + \lambda_{\mathrm{rn}} \kappa^* \big)\, \Delta t \Big], \qquad \kappa^* = \int (e^y - 1)\, P_{J,\mathrm{rn}}(y)\, dy, \label{rn_normratio}
\end{equation}
with $\kappa^*$ the risk-neutral expected proportional jump size. Equating to $e^{r_f \Delta t}$ and taking logarithms reduces no-arbitrage to a single scalar equation in the single unknown $\theta$,
\begin{equation}
\mu_c + \theta \sigma^2 + \lambda_{\mathrm{rn}}(\theta)\, \kappa^*(\theta) = r_f \qquad \Longleftrightarrow \qquad \mu_{c,\mathrm{rn}} = r_f - \lambda_{\mathrm{rn}}\, \kappa^*. \label{driftcorrection}
\end{equation}
The left-hand side is strictly increasing in $\theta$,\footnote{From (\ref{esscher_intensity}) and the definition of $\kappa^*$ one has $\lambda_{\mathrm{rn}}\kappa^* = \lambda\,[M_J(\theta+1) - M_J(\theta)]$, so the left-hand side of (\ref{driftcorrection}) is $\mu_c + \theta\sigma^2 + \lambda[M_J(\theta+1) - M_J(\theta)]$, with derivative $\sigma^2 + \lambda \int y\,(e^y - 1)\,e^{\theta y} P_J(y)\,dy \ge \sigma^2 > 0$, since $y(e^y - 1) \ge 0$ for all $y$. Equivalently, $\ln Z$ is a convex cumulant generating function, so its first difference $\ln[Z(\theta+1)/Z(\theta)]$ increases in $\theta$.} so the root is unique: the log-return constraint selects the Esscher family, and the martingale condition fixes its unique member, one of the infinitely many equivalent martingale measures the incomplete market admits. The right-hand form displays the jump compensation $-\lambda_{\mathrm{rn}}\kappa^*$ that the continuous drift must carry for the discounted price to be a martingale, and reproduces the resolved identity (\ref{jensen_resolved}), as it must. For the Merton specification (\ref{mertonjump}) every quantity is explicit: the tilted jump density is again Gaussian, $P_{J,\mathrm{rn}} = \mathcal{N}(\kappa_{\mathrm{rn}}, \delta^2)$ with risk-neutral mean $\kappa_{\mathrm{rn}} = \kappa + \theta \delta^2$ and rescaled intensity $\lambda_{\mathrm{rn}} = \lambda\, e^{\theta \kappa + \frac{1}{2}\theta^2 \delta^2}$, and the expected proportional jump is
\begin{equation}
\kappa^* = \exp\!\big( \kappa_{\mathrm{rn}} + \tfrac{1}{2}\delta^2 \big) - 1 = \exp\!\big( \kappa + \theta \delta^2 + \tfrac{1}{2}\delta^2 \big) - 1, \label{kappastar_merton}
\end{equation}
so that (\ref{driftcorrection}) determines $\theta$ uniquely, to be obtained numerically; the equation is transcendental and has no solution in elementary form, but every risk-neutral parameter is an explicit function of its root. The risk-neutral transition density then retains the form (\ref{infinitesimal}) under the replacements $\mu_c \to \mu_{c,\mathrm{rn}}$, $\lambda \to \lambda_{\mathrm{rn}}$, $P_J \to P_{J,\mathrm{rn}}$.

That no-arbitrage is imposed through the mean log-return is a choice, and the incompleteness of the market is what gives it weight. Had it been imposed instead on the gross return, with $\langle e^x \rangle$ rather than $\langle x \rangle$ as the controlled observable, the same maximization would have returned a tilt $P_{\mathrm{rn}} \propto \exp(\eta\, e^x)\, P_{JD}$, exponential in $e^x$: the minimum-relative-entropy, or canonical, martingale measure \cite{Stutzer96, Frittelli00}, which is equally free of arbitrage but does not preserve the jump-diffusion family and yields no closed-form premium. The two coincide in the diffusion limit $\lambda \to 0$, where the equivalent martingale measure is unique, and diverge once $\lambda > 0$. This gives the entropic framework its reading of incompleteness: \emph{the non-uniqueness of the equivalent martingale measure is the non-uniqueness of the informational constraint through which no-arbitrage is imposed}, each admissible controlled observable generating one member of the family, the mean log-return the Esscher measure and the gross return the canonical one. We adopt the log return because it is the observable on which the dynamics is already built, so that no new dynamical observable enters beyond those already governing the dynamics.

With the pricing measure in hand, the option is valued by the principle of risk-neutral valuation: its premium is the payoff it will deliver at maturity, taken in expectation under $P_{\mathrm{rn}}$ and discounted to the present at the risk-free rate. For a European call of strike $K$ and maturity $T$,

\begin{equation}
C = e^{-r_f T} \, \left\langle (S_T - K)^+ \,\big|\, S_0 \right\rangle_{\mathrm{rn}}, \label{mertonprice}
\end{equation}
where $(S_T - K)^+ = \max(S_T - K, 0)$ is the terminal payoff of the call, the expectation $\langle \cdot \rangle_{\mathrm{rn}}$ runs over the risk-neutral distribution of the terminal price $S_T$ given the present price $S_0$, and the factor $e^{-r_f T}$ carries that expected payoff back to today.

To make this explicit over a finite horizon $[0, T]$ with uniform parameters, the count of jumps is Poisson with mean $\lambda_{\mathrm{rn}} T$, accumulated from the Bernoulli short step in \ref{app:poisson_finite}; conditional on exactly $n$ jumps the terminal log price is Gaussian, its mean shifted by $n \kappa_{\mathrm{rn}}$ and its variance broadened by $n \delta^2$. The risk-neutral density of the terminal price is therefore a Poisson-weighted mixture of lognormals,
\begin{multline}
P_{\mathrm{rn}}(\ln S_T \, | \, \ln S_0) = \sum_{n=0}^{\infty} \frac{(\lambda_{\mathrm{rn}} T)^n \, e^{-\lambda_{\mathrm{rn}} T}}{n!} \\
\times \, \mathcal{N}\!\left( \ln S_0 + \big(r_f - \lambda_{\mathrm{rn}} \kappa^* - \tfrac{1}{2} \sigma^2\big) T + n \kappa_{\mathrm{rn}}, \; \sigma^2 T + n \delta^2 \right), \label{rn_mixture}
\end{multline}
with $\kappa_{\mathrm{rn}} = \kappa + \theta \delta^2$ the risk-neutral jump mean. Integrating the discounted payoff (\ref{mertonprice}) over this mixture branch by branch, each Gaussian branch contributing a Black--Scholes price, collapses the premium to Merton's closed-form valuation,
\begin{equation}
C = \sum_{n=0}^{\infty} \frac{(\lambda_{\mathrm{rn}} T)^n \, e^{-\lambda_{\mathrm{rn}} T}}{n!} \, C_{\mathrm{BS}}\!\left( S_0, K, T; \, r_n, \sigma_n^2 \right), \label{merton_price}
\end{equation}
the Poisson-weighted sum of Black--Scholes prices originally obtained by Merton \cite{Merton76jump}, with effective variance $\sigma_n^2 = \sigma^2 + n \delta^2 / T$ and effective rate $r_n = r_f - \lambda_{\mathrm{rn}} \kappa^* + n(\kappa_{\mathrm{rn}} + \tfrac{1}{2}\delta^2)/T$. Put--call parity continues to hold by the same model-independent arbitrage argument.

Setting $\lambda = 0$ in the general jump-diffusion derivation above degenerates the Esscher tilt: the rescaled jump intensity vanishes, $\lambda_{\mathrm{rn}} = 0$, and the drift correction $\lambda_{\mathrm{rn}} \kappa^*$ vanishes. The continuous-channel drift becomes simply

\begin{equation}
\mu_c = r_f, \label{rnc}
\end{equation}
recovering the classical Black--Scholes risk-neutrality constraint \cite{Hull}. The risk-neutral transition density on the no-jump branch reduces to the lognormal

\begin{equation}
P(\ln S^\prime \, | \, \ln S) = \frac{1}{Z} \, \exp  \left[ -\frac{1}{2 \sigma^2 \, \Delta t} \left( \ln S^\prime - \big(\ln S + r_f \Delta t - \tfrac{1}{2} \sigma^2 \Delta t \big) \right)^2 \right], \label{BSrn}
\end{equation}
and for a finite maturity $T$ with uniform parameters,

\begin{align}
P(S_T \, | \, S_0) \sim \mathrm{LN}\!\left( \ln S_0 + r_f \, T - \tfrac{1}{2} \sigma^2 \, T, \, \sigma \sqrt{T} \right),
\end{align}
a lognormal distribution on the terminal price. The European call value is the discounted expected payoff

\begin{equation}
C = e^{-r_f T} \int_K^\infty dS \, P(S, T \, | \, S_0) \, (S - K),
\end{equation}
which evaluates by the standard Black--Scholes calculation to

\begin{equation}
C = S_0 \, N(d_1) - e^{-r_f T} K \, N(d_2), \label{BScall}
\end{equation}
where

\begin{equation}
d_1 = \frac{\ln(S_0/K) + r_f T + \tfrac{1}{2} \sigma^2 T}{\sigma \sqrt{T}}, \qquad d_2 = d_1 - \sigma \sqrt{T},
\end{equation}
and $N(\cdot)$ is the standard normal cumulative distribution function. The corresponding put premium is

\begin{equation}
P_{\text{put}} = e^{-r_f T} K \, N(-d_2) - S_0 \, N(-d_1), \label{BSput}
\end{equation}
and the two satisfy put--call parity

\begin{equation}
C - P_{\text{put}} = e^{-r_f T} \, (F - K), \qquad S_0 = e^{-r_f T} F, \label{parity}
\end{equation}
where $F$ is the forward price. The Black--Scholes risk-neutral measure and the call/put pricing formulas are therefore the no-jump limit of the Maximum Entropy risk-neutral derivation above: they correspond to the corner case $\lambda = 0$ of the general jump-diffusion framework derived in this subsection.


\subsection{The Merton PIDE for the Option Premium}\label{subsec:MertonPIDE}
Having obtained the premium as a discounted expectation, we now ask how it evolves in time. The answer is a partial integro-differential equation (PIDE) for the premium, the option-pricing counterpart of the forward Kolmogorov--Feller equation derived for the density in Section~\ref{subsec:KF}. The derivation parallels that of the Black--Scholes--Merton equation in the continuous-only model, the jump channel contributing one further, integral term; \emph{Merton's PIDE} is the result, and the Black--Scholes--Merton (BSM) equation reappears as its no-jump limit $\lambda_{\mathrm{rn}} = 0$. We begin with the undiscounted expected payoff under the risk-neutral measure,

\begin{equation}
V( \ln S, K , t) = \int d \ln S_T \, P_{\mathrm{rn}}(\ln S_T, T \,|\, \ln S , t) \, \left( S_T - K \right), \label{K}
\end{equation}
where $P_{\mathrm{rn}}$ is the risk-neutral transition density of Section~\ref{subsec:Esscher}, the integration limits are left implicit so that the same expression serves the call and the put, and the discount factor is restored at the end. Differentiating with respect to time and carrying the derivative under the integral,

\begin{equation}
\partial_t V = \int d \ln S_T \, ( S_T - K) \, \partial_t P_{\mathrm{rn}}(\ln S_T, T \,|\, \ln S , t),  \label{V}
\end{equation}
the time dependence is carried entirely by the transition density, whose evolution in the \emph{backward}, or conditioning, argument is governed by the Backward Kolmogorov--Feller equation, the adjoint of the forward equation (\ref{KF}) under the risk-neutral measure,

\begin{align}
\partial_t P_{\mathrm{rn}} = & -\!\left( r_f - \tfrac{1}{2} \sigma^2 - \lambda_{\mathrm{rn}} \, \kappa^* \right) \frac{\partial P_{\mathrm{rn}}}{\partial \ln S} - \frac{\sigma^2}{2} \frac{\partial^2 P_{\mathrm{rn}}}{\partial (\ln S)^2} \nonumber \\
& - \lambda_{\mathrm{rn}} \!\int dy \, \big[ P_{\mathrm{rn}}(\ln S_T, T \,|\, \ln S + y, t) - P_{\mathrm{rn}}(\ln S_T, T \,|\, \ln S, t) \big] \, P_{J,\mathrm{rn}}(y), \label{BKF}
\end{align}
in which every derivative and shift acts on the conditioning argument $\ln S$, the forward argument $(\ln S_T, T)$ being held fixed throughout; in the integral term the conditioning argument is displaced to $\ln S + y$, so that $P_{\mathrm{rn}}(\ln S_T, T \,|\, \ln S + y, t)$ is the same terminal point reached from the shifted starting point $\ln S + y$. The drift coefficient $r_f - \tfrac{1}{2} \sigma^2 - \lambda_{\mathrm{rn}} \kappa^*$ carries the jump compensation (\ref{driftcorrection}) that renders the discounted underlying a martingale under $P_{\mathrm{rn}}$. Inserting (\ref{BKF}) into (\ref{V}) and using the linearity of the integral against the payoff $(S_T - K)$ transfers the backward operator from the density to the premium itself,

\begin{align}
\partial_t V = & -\!\left( r_f - \tfrac{1}{2} \sigma^2 - \lambda_{\mathrm{rn}} \, \kappa^* \right) \frac{\partial V}{\partial \ln S} - \frac{\sigma^2}{2} \frac{\partial^2 V}{\partial (\ln S)^2} \nonumber \\
& - \lambda_{\mathrm{rn}} \!\int dy \, \big[ V(\ln S + y, t) - V(\ln S, t) \big] \, P_{J,\mathrm{rn}}(y).
\end{align}
Recast in the price $S$ rather than its logarithm, the equation reads

\begin{equation}
\partial_t V + (r_f - \lambda_{\mathrm{rn}} \kappa^*) \, S \, \frac{\partial V}{\partial S} + \frac{\sigma^2 \, S^2}{2} \, \frac{\partial^2 V}{\partial S^2} + \lambda_{\mathrm{rn}} \!\int \big[ V(S e^y, t) - V(S, t) \big] \, P_{J,\mathrm{rn}}(y) \, dy = 0.
\end{equation}
The discount factor is restored through the substitution $E = e^{-r_f (T - t)} \, V$, which yields the equation for the premium itself,

\begin{equation}
\partial_t E + (r_f - \lambda_{\mathrm{rn}} \kappa^*) \, S \, \frac{\partial E}{\partial S} + \frac{1}{2} \sigma^2 \, S^2 \, \frac{\partial^2 E}{\partial S^2} + \lambda_{\mathrm{rn}} \!\int \big[ E(S e^y, t) - E(S, t) \big] \, P_{J,\mathrm{rn}}(y) \, dy - r_f \, E = 0, \label{MertonPIDE}
\end{equation}
where $E$ denotes the call and the put alike. This is \emph{Merton's option-pricing partial integro-differential equation} \cite{Merton76jump} for European options under jump-diffusion dynamics. Its structure mirrors the two channels of the underlying: the first three terms form the Black--Scholes--Merton differential operator, carrying the continuous contribution; the integral is the \emph{jump operator}, gathering the contributions of upward and downward jumps of size $y$ weighted by the risk-neutral jump-size density $P_{J,\mathrm{rn}}$; and the term $-r_f E$ effects the discounting. As in the dynamics, the compensator $-\lambda_{\mathrm{rn}} \kappa^*$ in the drift is precisely what holds the discounted underlying a martingale under $P_{\mathrm{rn}}$.

Setting $\lambda_{\mathrm{rn}} = 0$ silences both the integral term and the compensator, and the equation collapses to the celebrated Black--Scholes--Merton equation

\begin{equation}
\partial_t E + r_f \, S \, \frac{\partial E}{\partial S} + \frac{1}{2} \sigma^2 \, S^2 \, \frac{\partial^2 E}{\partial S^2} - r_f \, E = 0 \label{BSM}
\end{equation}
of the original continuous-only stock model. The PIDE (\ref{MertonPIDE}) is solved subject to the terminal condition $E(S, T) = (S - K)^+$ for the call or $E(S, T) = (K - S)^+$ for the put; for the Merton specification (\ref{mertonjump}) the solution is the closed-form Poisson-weighted sum of Black--Scholes prices.

\subsection{Implied Volatility Smile and Smirk}\label{subsec:smile}
The Black--Scholes model ascribes a single constant volatility $\sigma$ to the underlying log-returns, and so predicts that all European options on a given underlying imply the same volatility, whatever their strike $K$ or maturity $T$. The market does not comply. Implied volatility varies systematically with strike, tracing the celebrated \emph{volatility smile} or, in the equity case where it falls monotonically with strike, the \emph{volatility smirk}. Among the most robust regularities in all of derivatives markets, the pattern has been documented without interruption since the crash of 1987 \cite{Cont01stylized, Bates96}, and to reproduce it from first principles is a chief motivation for the jump extension developed here.

The market does not quote option prices directly but the single Black--Scholes volatility that reproduces them. For a call of strike $K$ and maturity $T$, the \emph{implied volatility} $\sigma_{\mathrm{imp}}(K, T)$ is the volatility at which the one-lognormal Black--Scholes formula returns the given premium; read against the entropic price (\ref{merton_price}), it is defined implicitly by
\begin{equation}
C_{\mathrm{BS}}\big( S_0, K, T;\, r_f,\, \sigma_{\mathrm{imp}}^{2}(K, T) \big) = \sum_{n=0}^{\infty} \frac{(\lambda_{\mathrm{rn}} T)^n \, e^{-\lambda_{\mathrm{rn}} T}}{n!} \, C_{\mathrm{BS}}\big( S_0, K, T;\, r_n, \sigma_n^{2} \big). \label{iv_def}
\end{equation}
The inversion is well posed: the Black--Scholes premium is strictly increasing in its volatility argument, so (\ref{iv_def}) fixes $\sigma_{\mathrm{imp}}$ uniquely at each $(K, T)$. The definition imports no assumption of its own; it is a pure change of variable, translating a price into the volatility the single-lognormal formula would require to produce it, so that the whole content of the smile resides in how that volatility must move with the strike. In the no-jump limit $\lambda_{\mathrm{rn}} = 0$ the sum collapses to its $n = 0$ term, the right-hand side is itself a Black--Scholes price at volatility $\sigma$, and (\ref{iv_def}) returns $\sigma_{\mathrm{imp}} = \sigma$ at every strike: the surface is flat, in plain contradiction with the data. Once jumps are admitted the right-hand side is a genuine Poisson-weighted mixture of lognormals of differing variances (\ref{rn_mixture}), which is not itself a lognormal, so no single volatility can reproduce it across strikes and $\sigma_{\mathrm{imp}}$ is forced to bend with $K$.

The implied-volatility surface read from (\ref{merton_price}) is no longer flat; its shape is the imprint of the mixture's departure from the single lognormal, a departure governed entirely by the three jump parameters. We make that dependence precise by computing the skewness and excess kurtosis of the risk-neutral terminal return.

Under the risk-neutral measure the terminal log-return $X = \ln(S_T/S_0)$ is the sum of a continuous Gaussian part and a compound-Poisson jump part,
\begin{equation}
X = \Big( r_f - \lambda_{\mathrm{rn}} \kappa^{*} - \tfrac{1}{2}\sigma^{2} \Big) T + \sigma\, W_T + \sum_{i=1}^{N} Y_i ,
\label{cum_X}
\end{equation}
where $W_T \sim \mathcal{N}(0,T)$, the count $N \sim \mathrm{Poisson}(\lambda_{\mathrm{rn}} T)$, and the jump sizes $Y_i \sim P_{J,\mathrm{rn}} = \mathcal{N}(\kappa_{\mathrm{rn}}, \delta^{2})$ are mutually independent and independent of $W_T$, with risk-neutral jump mean $\kappa_{\mathrm{rn}} = \kappa + \theta \delta^{2}$. The deterministic drift shifts only the mean and is immaterial to the standardized moments. Since the cumulant generating function of a sum of independent variables is the sum of theirs, the cumulants of $X$ add channel by channel: the Gaussian part contributes to the second cumulant alone, while the compound-Poisson part $J = \sum_i Y_i$ carries the cumulant generating function
\begin{equation}
K_J(u) = \lambda_{\mathrm{rn}} T \big( M_Y(u) - 1 \big), \qquad M_Y(u) = \langle e^{uY} \rangle ,
\label{cum_KJ}
\end{equation}
whose $k$-th derivative at the origin is $\lambda_{\mathrm{rn}} T \langle Y^{k}\rangle$. Collecting the two contributions,
\begin{align}
\kappa_2(X) &= \sigma^{2} T + \lambda_{\mathrm{rn}} T \,\langle Y^{2}\rangle , \label{cum_c2}\\
\kappa_3(X) &= \lambda_{\mathrm{rn}} T \,\langle Y^{3}\rangle , \label{cum_c3}\\
\kappa_4(X) &= \lambda_{\mathrm{rn}} T \,\langle Y^{4}\rangle . \label{cum_c4}
\end{align}
The diffusion enters the variance alone; every cumulant beyond the second is a pure jump effect, for a Gaussian has none. The standardized skewness and excess kurtosis,
\begin{equation}
\begin{aligned}
\mathrm{Skew}(X) &= \frac{\kappa_3(X)}{\kappa_2(X)^{3/2}}
= \frac{\lambda_{\mathrm{rn}} T \,\langle Y^{3}\rangle}{\big( \sigma^{2} T + \lambda_{\mathrm{rn}} T \,\langle Y^{2}\rangle \big)^{3/2}} , \\[4pt]
\mathrm{ExKurt}(X) &= \frac{\kappa_4(X)}{\kappa_2(X)^{2}}
= \frac{\lambda_{\mathrm{rn}} T \,\langle Y^{4}\rangle}{\big( \sigma^{2} T + \lambda_{\mathrm{rn}} T \,\langle Y^{2}\rangle \big)^{2}} ,
\end{aligned}
\label{cum_skewkurt}
\end{equation}
therefore originate entirely in the jump channel. For the Merton Gaussian jump distribution the raw moments are
\begin{equation}
\langle Y^{2}\rangle = \kappa_{\mathrm{rn}}^{2} + \delta^{2}, \qquad
\langle Y^{3}\rangle = \kappa_{\mathrm{rn}}\big( \kappa_{\mathrm{rn}}^{2} + 3\delta^{2} \big), \qquad
\langle Y^{4}\rangle = \kappa_{\mathrm{rn}}^{4} + 6\kappa_{\mathrm{rn}}^{2}\delta^{2} + 3\delta^{4},
\label{cum_moments}
\end{equation}
so both standardized cumulants are explicit functions of $(\lambda_{\mathrm{rn}}, \kappa_{\mathrm{rn}}, \delta, \sigma, T)$.

Three properties follow at once. First, $\langle Y^{4}\rangle$ is strictly positive, so $\mathrm{ExKurt}(X) > 0$ whenever $\lambda_{\mathrm{rn}} > 0$: the risk-neutral density is invariably leptokurtic, more sharply peaked and heavier-tailed than the Gaussian of equal variance. Second, the sign of the skewness is that of $\langle Y^{3}\rangle = \kappa_{\mathrm{rn}}(\kappa_{\mathrm{rn}}^{2} + 3\delta^{2})$, hence that of $\kappa_{\mathrm{rn}}$: downward mean jumps ($\kappa_{\mathrm{rn}} < 0$), the empirically dominant equity case, produce negative skewness, raising the value of out-of-the-money puts and bending the smile into a smirk. Third, at $\lambda_{\mathrm{rn}} = 0$ both quantities vanish, the density collapses to the single Black--Scholes lognormal, and the surface is flat. With the jump parameters held fixed, $\kappa_2$, $\kappa_3$, and $\kappa_4$ are each proportional to $T$, so in the regime of infrequent jumps $\lambda_{\mathrm{rn}} T \ll \sigma^{2} T/\langle Y^{2}\rangle$,
\begin{equation}
\mathrm{Skew}(X) \sim T^{-1/2}, \qquad \mathrm{ExKurt}(X) \sim T^{-1},
\label{cum_termstructure}
\end{equation}
the non-Gaussianity, and with it the steepness of the smile, decaying with maturity, in accord with the observed flattening of the surface at long horizons.

The link between these cumulants and the geometry of the surface can be displayed in closed form. Writing $\sigma_{\mathrm{ATM}}$ for the at-the-money implied volatility, about which the expansion is centred, and $d = [\ln(K/S_0) - (r_f - \tfrac{1}{2}\sigma_{\mathrm{ATM}}^{2})T] / (\sigma_{\mathrm{ATM}}\sqrt{T})$ for the standardized log-moneyness, an expansion of the Black--Scholes implied volatility in the standardized cumulants gives, to leading order,
\begin{equation}
\sigma_{\mathrm{imp}}(K,T) \approx \sigma_{\mathrm{ATM}} \bigg[ 1 + \tfrac{1}{6}\,\mathrm{Skew}(X)\, d + \tfrac{1}{24}\,\mathrm{ExKurt}(X)\, (d^{2} - 1) \bigg] .
\label{cum_ivexpansion}
\end{equation}
The two correction terms govern complementary features of the surface, and the model's regimes follow from the signs of the cumulants. The quadratic excess-kurtosis term $\tfrac{1}{24}\,\mathrm{ExKurt}\,(d^{2}-1)$ is positive in both deep wings ($|d| > 1$) and mildly negative near the money; acting alone, when the mean jump vanishes ($\kappa_{\mathrm{rn}} = 0$) while its variance does not, it raises both tails above the at-the-money level and produces a symmetric \emph{smile}, the signature of heavy tails without directional bias. The linear skewness term $\tfrac{1}{6}\,\mathrm{Skew}\,d$ breaks this symmetry. When the mean jump is downward ($\kappa_{\mathrm{rn}} < 0$), as it is for equities, the skewness is negative, the surface tilts down at high strikes and up at low, and the implied volatility falls monotonically with strike: this is the \emph{smirk}, the dominant equity pattern, steepened by a more negative $\kappa_{\mathrm{rn}}$ and deepened at the wings by a larger $\delta$. An upward mean jump ($\kappa_{\mathrm{rn}} > 0$) reverses the tilt and yields the \emph{reverse skew} of certain commodity and energy markets, where upside shocks outweigh downside. In the no-jump limit ($\lambda_{\mathrm{rn}} = 0$), finally, both cumulants vanish and $\sigma_{\mathrm{imp}} = \sigma_{\mathrm{ATM}}$ at every strike, recovering the flat Black--Scholes surface. A \emph{volatility frown}, by contrast, with implied volatility decreasing away from the money, is excluded from the outset: it would demand a platykurtic density, $\mathrm{ExKurt} < 0$, whereas the jump channel only ever injects probability mass into the tails and so raises the excess kurtosis above zero.

The branches with $n \geq 2$ are not negligible: for $\lambda_{\mathrm{rn}} T \approx 1$ they carry roughly a quarter of the probability mass and dominate the deep out-of-the-money wing, so the full sum (\ref{merton_price}), rather than a low-order truncation, is needed to reproduce the smirk at deep strikes.

The picture that emerges is transparent. The excess kurtosis, always positive, sets the curvature of the surface; the skewness, its sign that of $\kappa_{\mathrm{rn}}$, sets the tilt; and their magnitudes, fixed by the three jump parameters $(\lambda_{\mathrm{rn}}, \kappa_{\mathrm{rn}}, \delta)$, set the quantitative shape. To reproduce the volatility smile by appending a single arrival constraint and two jump-size moment constraints to the inference is the principal empirical achievement of the jump extension developed here: where the Black--Scholes--Merton model leaves the smile unexplained, the entropic framework delivers it, to leading order, as a derived consequence of the very constraints that fix the underlying dynamics.

\section{Summary and Discussion}\label{sec:summary}
We have set out an entropic framework for the dynamics of stocks and the pricing of European options in the presence of jumps. The dynamics is not postulated but derived: it is the distribution of greatest relative entropy on a chosen microstate, consistent with a small number of constraints and otherwise maximally noncommittal. This is the framework's principal contribution, and it stands in contrast to the prevailing practice of positing a stochastic process at the outset.

The microstate is the next log price, together with the number of jump arrivals over the interval and the individual jump sizes. Return symmetry selects the logarithm of price as the dynamical variable, and five pieces of information then fix the dynamics: continuity of motion and directionality on the continuous channel, and the arrival rate together with the first two moments of the jump size on the jump channel. Because these constraints act on disjoint parts of the microstate, the joint Maximum Entropy distribution factorizes into channel-specific factors, so that the independence of diffusion and jumps is a theorem of the inference rather than an assumption. The resulting process is the Merton jump-diffusion \cite{Merton76jump}, with Geometric Brownian Motion recovered when the jump intensity vanishes, and the log-price density obeys a Kolmogorov--Feller equation that reduces to the Fokker--Planck equation in the same limit.

The same principle prices the options. Because the jump market is incomplete its equivalent martingale measures are infinitely many, and no-arbitrage alone selects none; imposing it through the mean log-return, the observable that already governs the dynamics, and maximizing the relative entropy from the jump-diffusion measure selects the Esscher transform \cite{GerberShiu}, here derived rather than imported from actuarial science. A different controlled observable would yield a different measure, the gross return giving the canonical, minimum-relative-entropy one \cite{Stutzer96, Frittelli00}, so that the non-uniqueness of the pricing measure is that of the constraint through which no-arbitrage is imposed. The premium obeys Merton's partial integro-differential equation, and the classical Black--Scholes measure, the call and put formulas, and the Black--Scholes--Merton equation all return in the no-jump limit. The risk-neutral terminal density is a Poisson-weighted mixture of lognormals whose skewness and excess kurtosis, both of purely jump origin, generate the implied-volatility smile and, for downward jumps, the equity smirk.

What changes from one model to the next is never the inference but the information supplied to it. The framework is in this sense \emph{constraint-modular}: a new model is obtained by adding or relaxing a constraint, not by postulating a new process. A different choice of jump-size constraints reaches other members of the L\'evy family \cite{ContTankov04}, among them the asymmetric double-exponential model of Kou \cite{Kou02jump} and the variance gamma \cite{MadanCarrChang98}, normal inverse Gaussian \cite{BarndorffNielsen97}, and CGMY \cite{CGMY02} processes; the transition density, the Kolmogorov--Feller equation, and Merton's PIDE all carry over unchanged under the corresponding replacement of the jump-size distribution.

A deeper observation concerns the relationship between the two channels. Conditioned on $n$ jumps the log return is itself Gaussian (\ref{conditional_n}), so that removing the continuous channel, $\sigma \to 0$, leaves a pure-jump process whose returns remain Gaussian. Conversely, in the regime of frequent but vanishingly small jumps, $\lambda \to \infty$ with $\kappa \to 0$ and $\delta \to 0$ at fixed $\lambda\kappa$ and $\lambda\delta^2$, the compound-Poisson process converges to a Wiener process, and diffusion is recovered as the accumulation of infinitely many infinitesimal jumps. Diffusion is therefore not a separate primitive but a limiting case of the jump channel; the distinction between the two is one of constraint regime, not of kind.

The process derived here is Markovian, each constraint being expressed relative to the present state. Memory enters precisely when the constraints are conditioned on functionals of the past price path, at which point the increments cease to be independent and the dynamics turns non-Markovian. This is the natural route to several extensions: relaxing the uniformity of drift and volatility, stochastic volatility in the manner of Heston and the stochastic-volatility-with-jumps model of Bates \cite{Heston93, Bates96}, jump clustering through self-exciting Hawkes intensities \cite{Hawkes71}, and the recently documented regimes of rough volatility \cite{GJR18rough} and path-dependent volatility \cite{GuyonLekeufack23}. The rough regime corresponds to an entropic clock that carries memory; the path-dependent regime, to a volatility constraint conditioned on the history of the price. These we leave to future work.

The framework also transfers to other securities and problems. For a foreign exchange rate, a companion work shows the construction to carry over directly, yielding the dynamics of the rate together with its Garman--Kohlhagen option model \cite{ABFX}; and the joint dynamics of many assets opens onto portfolio selection, where the entropic organization around relevant information invites a reworking of Markowitz's mean--variance theory able to absorb additional information, or an investor's private beliefs.

The most fundamental extension concerns the origin of the constraints themselves. Here they have been taken as given; yet a price is not an autonomous quantity but is set by the interplay of supply and demand. A deeper formulation would apply the entropic machinery not to the price but to the order flow, treating the arrival of buyers and sellers and the resulting price formation as the fundamental microstate, from which the constraints of the continuous and jump channels would emerge as effective descriptions in appropriate regimes. Such a construction would explain not only how the constraints adopted here arise but under what market conditions, placing the present model on a still firmer inferential foundation.

\vspace {6pt}

\noindent \textbf{Acknowledgments:} {The author is grateful to his colleagues at VanEck for many stimulating and insightful discussions.}

\vspace{6pt}

\noindent \textbf{Disclosure:}{ The views and opinions expressed are those of the author and are current as of the publication date, and are not necessarily those of VanEck or its employees. Nothing contained herein should be construed as investment advice. Certain statements contained herein may constitute projections, forecasts and other forward looking statements, which may not reflect actual results, are valid as of the date of this communication and subject to change without notice.}

\appendix
\renewcommand{\thesection}{Appendix~\Alph{section}}
\section{The Finite-Time Poisson Count}\label{app:poisson_finite}

Equation (\ref{rn_mixture}) weights the terminal density by the number of jumps over the finite horizon $[0, T]$, which it takes to be Poisson with mean $\lambda T$ (and $\lambda_{\mathrm{rn}} T$ under the risk-neutral measure). Here we derive that Poisson distribution \emph{directly from the Bernoulli short step} (\ref{bernoullicount}) on which the model is built, using only the Maximum Entropy procedure and a uniform prior, with no further assumption. The count is never taken to be Poisson; it is the accumulation of the binary short steps, and the Poisson distribution is the derived consequence.

\paragraph{The Bernoulli short step.}
Over a short step of length $\Delta t$, the binary count $x \in \{0,1\}$ under the arrival constraint $\langle x \rangle = \lambda \Delta t$, maximizing the relative entropy against the uniform reference on $\{0,1\}$, is the Bernoulli distribution (\ref{bernoullicount}),

\begin{equation}
P(x = 0) = 1 - \lambda \, \Delta t, \qquad P(x = 1) = \lambda \, \Delta t, \label{app:pf_bern}
\end{equation}
derived in Section~\ref{sec:constraints}. This is the unit from which the finite-time count is built.

\paragraph{The joint distribution over the interval.}
Partition $[0, T]$ into $N = T / \Delta t$ non-overlapping short steps, and let $x_i \in \{0,1\}$ be the count in step $i$, so the total number of jumps over the interval is

\begin{equation}
n = \sum_{i=1}^{N} x_i \in \{0, 1, \ldots, N\}. \label{app:pf_sum}
\end{equation}
The only information available is the arrival rate, one Bernoulli constraint per step, $\langle x_i \rangle = \lambda \Delta t$; nothing couples distinct steps. Maximizing the relative entropy of the joint distribution $P(x_1, \ldots, x_N)$ relative to the uniform reference $\omega = 1$ on $\{0,1\}^N$, subject to these $N$ constraints and normalization, gives the Gibbs form

\begin{equation}
P(x_1, \ldots, x_N) = \frac{1}{Z}\, e^{-\beta \sum_i x_i} = \prod_{i=1}^{N} \frac{e^{-\beta x_i}}{1 + e^{-\beta}}, \label{app:pf_joint}
\end{equation}
with a single multiplier $\beta$ shared across steps by the symmetry of the constraints. The exponent is additive, so the joint factorizes: the short steps are \emph{independent} as a derived consequence of the joint Maximum Entropy with uncoupled constraints, not a separate assumption. Each factor is the Bernoulli (\ref{app:pf_bern}), $e^{-\beta}/(1 + e^{-\beta}) = \lambda \Delta t$.

\paragraph{The total count is binomial.}
The probability that the total count (\ref{app:pf_sum}) equals $n$ is the sum of the joint (\ref{app:pf_joint}) over all configurations with exactly $n$ ones. Each such configuration carries the same weight $(\lambda \Delta t)^n (1 - \lambda \Delta t)^{N - n}$, and the number of them is the binomial coefficient $\binom{N}{n}$, so

\begin{equation}
P(n) = \binom{N}{n} (\lambda \Delta t)^n (1 - \lambda \Delta t)^{N - n}, \label{app:pf_binom}
\end{equation}
the binomial distribution. The factorial structure $1/n!$ of the Poisson distribution to come is already present here, inside $\binom{N}{n} = N! / [n! (N - n)!]$, as the multiplicity the uniform measure attaches to the macrostate $n$: it is counted, not inserted.

\paragraph{The continuum limit.}
Holding the horizon $T$ fixed and refining the step, $N \to \infty$ with $\lambda \Delta t = \lambda T / N$, we write (\ref{app:pf_binom}) as

\begin{equation}
P(n) = \frac{(\lambda T)^n}{n!} \cdot \underbrace{\frac{n!\,\binom{N}{n}}{N^n}}_{(\mathrm{i})} \cdot \underbrace{\Big( 1 - \tfrac{\lambda T}{N} \Big)^{N}}_{(\mathrm{ii})} \cdot \underbrace{\Big( 1 - \tfrac{\lambda T}{N} \Big)^{-n}}_{(\mathrm{iii})}.
\end{equation}
Factor (i) is $\prod_{k=0}^{n-1}(1 - k/N) \to 1$; factor (ii) $\to e^{-\lambda T}$ by the standard exponential limit; factor (iii) $\to 1$ for fixed $n$. Hence

\begin{equation}
\boxed{\; P(n) = \frac{(\lambda T)^n}{n!}\, e^{-\lambda T} \;} \label{app:pf_poisson}
\end{equation}
the Poisson distribution with mean $\lambda T$ used in (\ref{rn_mixture}); under the risk-neutral measure the identical argument with $\lambda \to \lambda_{\mathrm{rn}}$ gives Poisson$(\lambda_{\mathrm{rn}} T)$. Every step rests only on Maximum Entropy with a uniform prior: the Bernoulli short step is the MaxEnt output of the arrival constraint, its accumulation over the interval is the joint MaxEnt of uncoupled per-step constraints, the binomial total follows by enumeration, and the factorial $1/n!$ is the binomial multiplicity $\binom{N}{n} \sim N^n / n!$, not a chosen reference measure. No step is Poisson by assumption. Had the entropy instead been maximized directly on the bare count $n \in \mathbb{Z}_{\geq 0}$ with a uniform reference $\omega_n = 1$, the result would be the geometric distribution, not the Poisson; the Poisson is recovered precisely because the count is resolved into its underlying binary short steps, the factorial appearing as the multiplicity of the uniform measure on those steps.

\section{The Kolmogorov--Feller Equation}\label{app:KF}

In Section~\ref{subsec:KF} the forward equation (\ref{KF}) for the log-price density was obtained by taking the $\Delta t \to 0$ limit of the entropic-instant identity (\ref{CK}). Here we carry out that derivation in full. The starting point is the entropic instant,

\begin{equation}
p(\ln S, t + \Delta t) = \int d \ln S_0 \, P_{JD}(\ln S \, | \, \ln S_0) \, p(\ln S_0, t), \label{app:KF_CK}
\end{equation}
the propagation of the density over one step by the marginalized transition kernel (\ref{marginalized}), with $\ln S_0$ the log price at the start of the step. Writing the kernel in the displacement variable $y = \ln S - \ln S_0$ turns (\ref{app:KF_CK}) into the convolution

\begin{equation}
p(\ln S, t + \Delta t) = \int \! dy \, P_{JD}(\ln S \, | \, \ln S - y) \, p(\ln S - y, t), \label{app:KF_conv}
\end{equation}
where the transition kernel depends on its arguments only through the displacement $y$.

For an infinitesimal interval the kernel takes the two-channel form (\ref{infinitesimal}), in which a pure diffusion move occurs with probability $1 - \lambda \, \Delta t$ and a single jump superposed on the diffusion occurs with probability $\lambda \, \Delta t$, the multiple-jump contributions being $\mathcal{O}\big((\Delta t)^2\big)$. Substituting (\ref{infinitesimal}) into (\ref{app:KF_conv}),

\begin{align}
p(\ln S, t + \Delta t) &= (1 - \lambda \, \Delta t) \! \int \! dy \, P_c(y) \, p(\ln S - y, t) \nonumber \\
&\quad + \lambda \, \Delta t \! \int \! dy \, \big[\, P_c \, * \, P_J \,\big](y) \, p(\ln S - y, t) + \mathcal{O}\!\left((\Delta t)^2\right). \label{app:KF_split}
\end{align}
In the jump-channel term the diffusion factor may be replaced by a Dirac measure to leading order: the diffusion Gaussian $P_c$ has variance $\sigma^2 \Delta t$ and degenerates to $\delta$ as $\Delta t \to 0$, so $P_c * P_J \to P_J$ up to corrections that are higher order once multiplied by the prefactor $\lambda \, \Delta t$. This reduces (\ref{app:KF_split}) to

\begin{align}
p(\ln S, t + \Delta t) &= (1 - \lambda \, \Delta t) \! \int \! dy \, P_c(y) \, p(\ln S - y, t) \nonumber \\
&\quad + \lambda \, \Delta t \! \int \! dy \, P_J(y) \, p(\ln S - y, t) + \mathcal{O}\!\left((\Delta t)^2\right). \label{app:KF_infinitesimal}
\end{align}

We now expand the continuous-channel convolution to first order in $\Delta t$. The diffusion kernel $P_c$ of (\ref{phi_c}) is a Gaussian in $y$ with mean $(\mu_c - \tfrac{1}{2}\sigma^2)\, \Delta t$ and variance $\sigma^2 \Delta t$, both $\mathcal{O}(\Delta t)$. Taylor expanding $p(\ln S - y, t)$ about $y = 0$ and integrating term by term against $P_c$, only the first two moments survive at order $\Delta t$,

\begin{align}
& \int \! dy \, P_c(y) \, p(\ln S - y, t) \nonumber \\
& \quad = p(\ln S, t) - \langle y \rangle_{P_c} \, \frac{\partial p}{\partial \ln S} + \frac{1}{2} \langle y^2 \rangle_{P_c} \, \frac{\partial^2 p}{\partial (\ln S)^2} + \mathcal{O}\!\left( (\Delta t)^2 \right) \nonumber \\
& \quad = p(\ln S, t) - \big( \mu_c - \tfrac{1}{2} \sigma^2 \big) \, \Delta t \, \frac{\partial p}{\partial \ln S} + \frac{1}{2} \sigma^2 \, \Delta t \, \frac{\partial^2 p}{\partial (\ln S)^2} + \mathcal{O}\!\left( (\Delta t)^2 \right), \label{app:KF_diffexp}
\end{align}
where $\langle y \rangle_{P_c} = (\mu_c - \tfrac{1}{2}\sigma^2)\,\Delta t$ and $\langle y^2 \rangle_{P_c} = \sigma^2 \Delta t + \mathcal{O}\big((\Delta t)^2\big)$ are the mean and second moment of the diffusion kernel. Inserting (\ref{app:KF_diffexp}) into (\ref{app:KF_infinitesimal}) and discarding the $\lambda \, \Delta t$ multiple of the $\mathcal{O}(\Delta t)$ diffusion corrections, which are second order,

\begin{align}
p(\ln S, t + \Delta t) ={} & p(\ln S, t) - \big( \mu_c - \tfrac{1}{2} \sigma^2 \big) \, \Delta t \, \frac{\partial p}{\partial \ln S} + \frac{1}{2} \sigma^2 \, \Delta t \, \frac{\partial^2 p}{\partial (\ln S)^2} \nonumber \\
& + \lambda \, \Delta t \! \int \! dy \, \big[ p(\ln S - y, t) - p(\ln S, t) \big] \, P_J(y) + \mathcal{O}\!\left( (\Delta t)^2 \right), \label{app:KF_assembled}
\end{align}
where the $-\lambda \, \Delta t \, p(\ln S, t)$ piece needed to form the bracketed difference has been supplied by the $(1 - \lambda \, \Delta t)$ prefactor of the diffusion term, using $\int dy \, P_J(y) = 1$. Subtracting $p(\ln S, t)$ from both sides, dividing by $\Delta t$, and letting $\Delta t \to 0$,

\begin{equation}
\partial_t p(\ln S, t) = \lim_{\Delta t \to 0} \frac{p(\ln S, t + \Delta t) - p(\ln S, t)}{\Delta t}, \label{app:KF_deriv}
\end{equation}
yields the Kolmogorov--Feller equation

\begin{align}
\partial_t p(\ln S, t) = & -\frac{\partial}{\partial \ln S} \!\left[ \! \big( \mu_c - \tfrac{1}{2} \sigma^2 \big) \, p(\ln S, t) \! \right] + \frac{1}{2} \frac{\partial^2}{\partial (\ln S)^2} \!\left[ \sigma^2 \, p(\ln S, t) \right] \nonumber \\
& + \lambda \!\int \! dy \, \big[ p(\ln S - y, t) - p(\ln S, t) \big] \, P_J(y), \label{app:KF_result}
\end{align}
which is Equation (\ref{KF}) of Section~\ref{subsec:KF}. The constants $\mu_c$ and $\sigma^2$ are carried outside the derivatives because they are uniform parameters here; relaxing that assumption to space- or time-dependent coefficients leaves the derivation otherwise unchanged and reproduces the divergence-form diffusion operator displayed in (\ref{app:KF_result}).


\section{The Mean Log-Return from the Short-Step Increment}\label{app:mlr}

The resolved constraint (\ref{jensen_resolved}) was obtained in Section~\ref{subsec:Esscher} by expanding the logarithm under the pricing measure and retaining the jump compensator to all orders. Here we derive the same identity directly from the short-step price increment (\ref{jd_SDE}). The calculation makes transparent why the continuous channel truncates at second order in the manner of It\^o while the jump channel contributes at every order, the infinite series resumming to the additive log-jump that (\ref{jd_logSDE}) carries from the outset.

\paragraph{The increment and its orders.}
Over an infinitesimal step the price increment (\ref{jd_SDE}) is
\begin{equation}
\frac{\Delta S}{S} = \mu_c \, \Delta t + \Delta W + (e^{Y} - 1)\, \Delta N , \label{app:mlr_increment}
\end{equation}
the three contributions being of distinct character. The drift $\mu_c \Delta t$ is of order $\Delta t$; the Wiener increment $\Delta W$ is of order $(\Delta t)^{1/2}$, with $\langle \Delta W \rangle = 0$ and $\langle \Delta W^2 \rangle = \sigma^2 \Delta t$; and the jump $(e^{Y}-1)\Delta N$ is rare but finite, the Bernoulli increment $\Delta N \in \{0,1\}$ firing with probability $\lambda \Delta t$ and carrying a displacement of order unity. Because $\Delta N^{k} = \Delta N$ for a binary count, every power of the jump term is of the same order,
\begin{equation}
\big\langle [(e^{Y}-1)\,\Delta N]^{k} \big\rangle = \lambda \, \Delta t \, \big\langle (e^{Y}-1)^{k} \big\rangle = \mathcal{O}(\Delta t), \qquad k \ge 1 . \label{app:mlr_bern}
\end{equation}
The jump is therefore not a $(\Delta t)^{1/2}$ quantity that may be split off into a half-power and a full-power piece; it enters every power of the increment at the same order $\Delta t$, and this is exactly what sets it apart from the diffusion.

\paragraph{The moments of the increment.}
Writing $x = \Delta S / S$ and keeping terms to first order in $\Delta t$, the cross contributions vanish: the drift--diffusion product carries $\langle \Delta W \rangle = 0$, the drift--jump product is $\mathcal{O}\big((\Delta t)^2\big)$, and the diffusion--jump product factorizes through the independent $\langle \Delta W \rangle = 0$. The surviving moments are
\begin{align}
\langle x \rangle   &= \mu_c \, \Delta t + \lambda \, \Delta t \, \big\langle e^{Y} - 1 \big\rangle , \label{app:mlr_m1} \\
\langle x^{2} \rangle &= \sigma^{2} \Delta t + \lambda \, \Delta t \, \big\langle (e^{Y} - 1)^{2} \big\rangle , \label{app:mlr_m2} \\
\langle x^{k} \rangle &= \lambda \, \Delta t \, \big\langle (e^{Y} - 1)^{k} \big\rangle , \qquad k \ge 3 . \label{app:mlr_mk}
\end{align}
The diffusion enters the first and second moments alone; the jump enters every moment.

\paragraph{The logarithm, channel by channel.}
Expanding $\ln(S^\prime/S) = \ln(1 + x) = x - \tfrac{1}{2}x^{2} + \tfrac{1}{3}x^{3} - \cdots$ and taking the expectation term by term, the two channels separate cleanly. The continuous channel, present only in $\langle x \rangle$ and $\langle x^{2} \rangle$, contributes the It\^o drift
\begin{equation}
\mu_c \, \Delta t - \tfrac{1}{2}\sigma^{2}\,\Delta t , \label{app:mlr_diff}
\end{equation}
the term $-\tfrac{1}{2}\sigma^{2}\Delta t$ being the familiar second-order correction beyond which the continuous expansion does not reach. The jump channel, present in every $\langle x^{k} \rangle$, contributes the full series
\begin{equation}
\lambda \, \Delta t \, \Big\langle (e^{Y}-1) - \tfrac{1}{2}(e^{Y}-1)^{2} + \tfrac{1}{3}(e^{Y}-1)^{3} - \cdots \Big\rangle
= \lambda \, \Delta t \, \big\langle \ln\!\big( 1 + (e^{Y}-1) \big) \big\rangle
= \lambda \, \Delta t \, \langle Y \rangle , \label{app:mlr_jumpsum}
\end{equation}
the infinite sum collapsing to $\langle Y \rangle$ because $\ln\!\big(1 + (e^{Y}-1)\big) = \ln e^{Y} = Y$. This is precisely the additive log-jump that (\ref{jd_logSDE}) carries from the start: in the price variable the jump appears as the proportional displacement $e^{Y}-1$ and demands the whole Taylor series, whereas in the log variable it is simply $Y$.

\paragraph{The mean log-return.}
Collecting the two contributions,
\begin{equation}
\Big\langle \ln \frac{S^\prime}{S} \Big\rangle = \big( \mu_c - \tfrac{1}{2}\sigma^{2} \big)\Delta t + \lambda \, \langle Y \rangle \, \Delta t . \label{app:mlr_result}
\end{equation}
Under the pricing measure the parameters take their risk-neutral values, and the no-arbitrage drift condition (\ref{driftcorrection}), $\mu_{c,\mathrm{rn}} = r_f - \lambda_{\mathrm{rn}} \kappa^{*}$ with $\kappa^{*} = \langle e^{Y}-1 \rangle_{\mathrm{rn}}$, gives
\begin{align}
\Big\langle \ln \frac{S^\prime}{S} \Big\rangle_{\mathrm{rn}}
&= \big( \mu_{c,\mathrm{rn}} - \tfrac{1}{2}\sigma^{2} \big)\Delta t + \lambda_{\mathrm{rn}} \langle Y \rangle_{\mathrm{rn}} \, \Delta t \nonumber \\
&= \big( r_f - \tfrac{1}{2}\sigma^{2} \big)\Delta t - \lambda_{\mathrm{rn}} \big( \kappa^{*} - \langle Y \rangle_{\mathrm{rn}} \big)\Delta t \nonumber \\
&= \big( r_f - \tfrac{1}{2}\sigma^{2} \big)\Delta t - \lambda_{\mathrm{rn}} \big\langle e^{Y} - 1 - Y \big\rangle_{\mathrm{rn}} \, \Delta t , \label{app:mlr_rn}
\end{align}
since $\kappa^{*} - \langle Y \rangle_{\mathrm{rn}} = \langle e^{Y}-1 \rangle_{\mathrm{rn}} - \langle Y \rangle_{\mathrm{rn}} = \langle e^{Y}-1-Y \rangle_{\mathrm{rn}}$. The last line is the resolved identity (\ref{jensen_resolved}).

The contrast between the channels is the whole content of the calculation. The diffusion increment is of order $(\Delta t)^{1/2}$, so the expansion of the logarithm closes at second order and leaves the single It\^o correction $-\tfrac{1}{2}\sigma^{2}$. The jump increment is of order unity in size but $\mathcal{O}(\Delta t)$ in probability, so it feeds every order of the expansion at the same weight, and only the resummation of the entire series, to $\langle Y \rangle$, recovers the additive log-jump. The convexity term $-\lambda_{\mathrm{rn}}\langle e^{Y}-1-Y \rangle_{\mathrm{rn}}$ is the residue of that resummation, measuring the gap between the proportional jump $e^{Y}-1$ that drives the price and the additive jump $Y$ that drives its logarithm.


\end{document}